\documentclass[11pt]{article} %
\usepackage[T1]{fontenc}
\usepackage{amssymb, amsmath, amsthm, graphicx, subfigure, mathtools}
\usepackage[margin=1in]{geometry}
\mathtoolsset{showonlyrefs}
\usepackage{tikz}

\usepackage{float}
\usepackage[colorlinks=true, allcolors=blue]{hyperref}

\usepackage{xcolor}
\usepackage{braket}
\usepackage{esvect}
\usepackage{enumerate}
\usepackage{scalefnt}

\usepackage{algorithm}
\usepackage{algorithmicx}
\usepackage[noend]{algpseudocode}

\usepackage{pifont}%

\newtheorem{theorem}{Theorem}[section]
\newtheorem{definition}[theorem]{Definition}

\newtheorem{corollary}[theorem]{Corollary}
\newtheorem{lemma}[theorem]{Lemma}

\newtheorem{remark}[theorem]{Remark}

\newcommand{\bone}{\mathbf{1}}

\newcommand{\br}{\mathbf{r}}
\newcommand{\bu}{\mathbf{u}}
\newcommand{\bv}{\mathbf{v}}
\newcommand{\bx}{\mathbf{x}}
\newcommand{\by}{\mathbf{y}}

\newcommand{\RR}{\mathbb{R}}

\renewcommand{\Pr}{\mathbb{P}}

\newcommand{\one}{\mathbf 1}

\newcommand{\E}{\mathop{{}\mathbb{E}}}

\newcommand{\GE}{\;\ge\;}
\newcommand{\LE}{\;\le\;}
\newcommand{\EQ}{\;=\;}

\newcommand{\LT}{\;<\;}

\newcommand{\comp}{\mathsf{Completeness}}
\newcommand{\sound}{\mathsf{Soundness}}

\newcommand{\THRESH}{{\cal THRESH}}

\newcommand{\agw}{\alpha_{GW}}
\newcommand{\bgw}{b_{GW}}
\newcommand{\cgw}{c_{GW}}

\newcommand{\SDP}{\operatorname{SDP}}
\newcommand{\OPT}{\operatorname{OPT}}

\newcommand{\eps}{\varepsilon}

\numberwithin{equation}{section}

\begin{document}

\title{MAX BISECTION might be harder\\ to approximate than MAX CUT}

\date{}

\author{Joshua Brakensiek\thanks{University of California, Berkeley. Email: \texttt{josh.brakensiek@berkeley.edu}. Supported in part by a Simons Investigator award and NSF grants CCF-2211972 and DMS-2503280.}\and Neng Huang\thanks{University of Michigan. Email: \texttt{nengh@umich.edu}}\and Aaron Potechin\thanks{University of Chicago. Email: \texttt{potechin@uchicago.edu}}\and Uri Zwick\thanks{Blavatnik School of Computer Science, Tel Aviv University, Israel. Email: \texttt{zwick@tau.ac.il}. Work supported by ISF grant 2735/2025.}}

\maketitle

\begin{abstract}
\setlength{\parindent}{0pt}
\setlength{\parskip}{6pt plus 2pt}
\noindent
The MAX BISECTION problem seeks a maximum-size cut that evenly divides the vertices of a given undirected graph. An open problem raised by Austrin, Benabbas, and Georgiou [SODA'13, TALG'16] is whether MAX BISECTION can be approximated as well as MAX CUT, i.e., to within ${\alpha_{\text{GW}}}\approx 0.8785672\ldots$, which is the approximation ratio achieved by the celebrated Goemans-Williamson algorithm for MAX CUT, which is best possible assuming the Unique Games Conjecture (UGC). They conjectured that the answer is yes.

The current paradigm for obtaining approximation algorithms for MAX BISECTION, due to Raghavendra and Tan [SODA'12] and Austrin, Benabbas, and Georgiou, follows a two-phase approach. First, a large number of rounds of the Sum-of-Squares (SoS) hierarchy is used to find a solution to the ``Basic SDP'' relaxation of MAX CUT which is \emph{$\varepsilon$-uncorrelated}, for an arbitrarily small $\varepsilon > 0$. Second, standard SDP rounding techniques (such as ${\cal THRESH}$) are used to round this $\varepsilon$-uncorrelated solution, producing with high probability a cut that is almost balanced, i.e., a cut that has at most $\frac12+\varepsilon$ fraction of the vertices on each side. This cut is then converted into an exact bisection of the graph with only a small loss.

In this paper, we show that this two-stage paradigm cannot be used to obtain an $\alpha_{GW}$-approximation algorithm for MAX BISECTION if one relies only on the $\varepsilon$-uncorrelatedness property of the solution produced by the first phase. More precisely, for any $\varepsilon > 0$, we construct an explicit instance of MAX BISECTION for which the ratio between the value of the optimal integral solution and the value of some $\varepsilon$-uncorrelated solution of the Basic SDP relaxation is less than $0.87853 < {\alpha_{\text{GW}}}$. Our instances are also integrality gaps for the Basic SDP relaxation of MAX BISECTION. As a corollary, we also obtain a family of MAX BISECTION-based dictatorship tests (in the sense of Raghavendra and Tan) with the same ratio.
\end{abstract}

\setcounter{page}{0}\thispagestyle{empty}\clearpage

\pagebreak

\section{Introduction}

MAX CUT is one of the most natural \emph{Maximum Constraint Satisfaction Problems (MAX CSPs)}. (See, e.g., \cite{KSTW01}.) The input is a weighted undirected graph $G=(V,E,w)$, where $w:E\to\RR^+$. The goal is to find a cut $(S,V\setminus S)$, where $S\subseteq V$, that maximizes the weight of the edges that cross the cut, i.e., $\sum_{u\in S, v\in V\setminus S} w(u,v)$. MAX CUT is one of the first NP-hard problems (Karp \cite{Karp72}), and it is also NP-hard to approximate to within a factor of $\frac{16}{17}+\varepsilon$, for any $\varepsilon>0$ (H{\aa}stad \cite{H01}). Goemans and Williamson \cite{goemans1995Improveda} obtained a $\alpha_{GW}$-approximation algorithm for MAX CUT using \emph{Semidefinite Programming (SDP)}, where $\alpha_{GW}=\min_{0< \theta\le \pi}\frac{2}{\pi}\frac{\theta}{1-\cos\theta}\approx 0.8785672\ldots$. Khot, Kindler, Mossel and O’Donnell \cite{KKMO07} and Mossel, O'Donnell, and Oleszkiewiczs~\cite{MOO10} showed, 
assuming the \emph{Unique Games Conjecture (UGC)} (Khot \cite{khot02}), that no polynomial-time $(\alpha_{GW}+\varepsilon)$-approximation algorithm for MAX CUT, for any $\varepsilon>0$, can be obtained.

MAX BISECTION is a natural variant of MAX CUT in which the cut $(S,V\setminus S)$ is required to be completely balanced, i.e., $|S|=|V\setminus S|=\frac{n}{2}$, where it is assumed that $n=|V|$ is even. This is perhaps the canonical example of a problem obtained by adding a \emph{global cardinality constraint} to a MAX CSP problem. There is a trivial approximation-preserving reduction from MAX CUT to MAX BISECTION, simply replace an input graph~$G$ by two independent copies of itself. Thus MAX BISECTION is at least as hard to approximate as MAX CUT. A natural question, raised in particular by Austrin, Benabbas, and Georgiou \cite{ABG16}, is whether MAX BISECTION can be approximated as well as MAX CUT.

Frieze and Jerrum \cite{FrJe97} obtained an $0.651$-approximation algorithm for MAX BISECTION, the first non-trivial approximation algorithm for MAX BISECTION, by modifying the MAX CUT approximation algorithm of Goemans and Williamson. They also say: ``The difficulty with generalizing Goemans and Williamson's heuristic to MAX BISECTION is that it does not generally give a bisection of $V$.'' Slightly improved approximation ratios were subsequently obtained by \cite{Ye01}, \cite{HaZw02} and by \cite{FL06}. 

Raghavendra and Tan \cite{RaTa12}, in a major breakthrough, showed how to use stronger SDP relaxations of MAX BISECTION, namely, using relaxations in the Lasserre Hierarchy \cite{Lasserre02}, also known as the \emph{Sum of Squares (SoS)} hierarchy, to obtain an $0.85$-approximation algorithm for MAX BISECTION. The stronger SDP relaxation helps with solving the intrinsic problem with the simpler SDP relaxation mentioned by Frieze and Jerrum, namely the lack of balance when using rounding procedures similar to those used by Goemans and Williamson. Austrin, Benabbas, and Georgiou \cite{ABG16} built on the results of Raghavendra and Tan and obtained a $0.8776$-approximation algorithm for MAX BISECTION.~\footnote{The analysis of this algorithm is computer-assisted. They also obtained a $0.8736$-approximation algorithm with an analytical proof.} This is very close to the presumably optimal $\alpha_{GW}\simeq 0.8785672$ approximation ratio of MAX CUT. Austrin, Benabbas and Georgiou \cite{ABG16} conjectured that there is an $(\alpha_{GW}-\varepsilon)$-approximation algorithm for MAX BISECTION, for any $\eps>0$, or in other words that MAX BISECTION and MAX CUT have the same approximation threshold.

In this paper, we give evidence that the conjecture of Austrin, Benabbas and Georgiou \cite{ABG16} might not be true. We do not refute the conjecture, but rather show that the framework used by Raghavendra and Tan \cite{RaTa12} and by Austrin, Benabbas, and Georgiou \cite{ABG16} does not give an approximation ratio of $0.87853$, which is strictly less than $\agw$. This implies that obtaining a better approximation ratio for MAX BISECTION (if it is even possible to do so) would require new algorithmic techniques. A more precise description of our results is given in Section~\ref{sub-summary}.
\subsection{The paradigm of Raghavendra-Tan and Austrin-Benabbas-Georgiou}\label{sec:rt_abg}

To properly describe our contributions, we need to first explain the paradigm used by Raghavendra and Tan~\cite{RaTa12} and by Austrin, Benabbas, and Georgiou~\cite{ABG16} to obtain the state-of-the-art approximation algorithms for MAX BISECTION. First, to set up some notation, we briefly recall the Goemans-Williamson (GW) algorithm~\cite{goemans1995Improveda} for MAX CUT. Consider a given graph $G = (V,E)$ where $V = [n] := \{1, 2, \hdots, n\}$. Let $\OPT_{CUT}(G)$ be the size of the largest cut of $G$. Using semidefinite programming (SDP), one can find unit vectors $\bv_1, \hdots, \bv_n \in \mathbb R^{n}$ which maximize $\SDP_{CUT}(G) := \sum_{\{i,j\} \in E} \frac{1 - \bv_i \cdot \bv_j}{2}$. \footnote{Throughout the paper $\bu \cdot \bv$ denotes the inner product of the vectors $\bu$ and $\bv$.}
Since any cut of $G$ can be encoded as a set of (one-dimensional) unit vectors, we have that $\SDP_{CUT}(G) \ge \OPT_{CUT}(G)$. To convert the SDP vectors into an actual cut, we use \emph{random hyperplane rounding}. Namely, we sample a random $n$-dimensional normal vector $\br \in \mathcal N(0, I_n)$. We partition $V$ into $(A,B)$ where $A := \{i \in V : \bv_i\cdot \br > 0\}$ and $B=V\setminus A$. 
A simple analysis shows that the probability that an edge $\{i,j\} \in E$ is cut by the cut $(A,B)$ is precisely $\frac{\cos^{-1}(\bv_i \cdot \bv_j)}{\pi}$. As the contribution of $\{i,j\} \in E$ to $\SDP_{CUT}(G)$ is $\frac{1-\bv_i \cdot \bv_j}{2}$, the approximation ratio achieved by the algorithm is thus
\[
    \agw \EQ \min_{b \in [-1,1)} \frac{\cos^{-1}(b)}{\pi} \cdot \frac{2}{1 - b} \;\approx\; 0.8785672\;,
\]
where the minimum is attained at $\bgw \approx -0.6891577$. 

Note that every vertex $i \in V$ is equally likely to be on either side of the cut. Despite this fact, the GW algorithm cannot be used directly as a MAX BISECTION algorithm. The issue is the \emph{correlations} between the vectors. As a hypothetical scenario, imagine there is some vector $\bv_0 \in \mathbb R^n$ such that $\bv_i\cdot\bv_0 \ge \frac12$ for all $i \in \bv_0$. In this case, whether some $i \in V$ lands in $A$ strongly correlates with whether $\bv_0\cdot \br > 0$. In particular, although the expected size of $A$ is $n/2$, the variance of the size of $A$ is $\Omega(n^2)$. 

Multiple adjustments were made by Raghavendra and Tan~\cite{RaTa12} and by Austrin, Benabbas, and Georgiou~\cite{ABG16} to circumvent this issue. First, as was also done in previous MAX BISECTION algorithms, they added an additional unit vector $\bv_0$ to the relaxation and imposed the constraint $\sum_{i=1}^n \bv_0 \cdot \bv_i = 0$. Here, we let $\bv_0 \cdot \bv_i$ be the \emph{bias} of $\bv_i$, where $\frac{1-\bv_0\cdot \bv_i}{2}$ %
is the probability the SDP (implicitly) assigns to $i$ being on the $A$-side of the cut. However, this constraint by itself does not change anything, as we can always pick $\bv_0$ to be orthogonal to every other vector. %

To make this constraint meaningful, the main innovation introduced by Raghavendra and Tan~\cite{RaTa12} and by Austrin, Benabbas, and Georgiou~\cite{ABG16}, was to make the vectors \emph{$\eps$-uncorrelated}. For all $i \in V$, let $\bv_i^{\perp} = \frac{\bv_i - (\bv_0 \cdot \bv_i) \bv_i}{\sqrt{1 - (\bv_0 \cdot \bv_i)^2}}$ be the normalized projection of $\bv_i$ on the space perpendicular to~$\bv_0$. We say unit vectors $\bv_0, \bv_1, \hdots, \bv_n \in \mathbb R^{n+1}$ are $\eps$-uncorrelated\footnote{We note that the definition of $\eps$-uncorrelated in Raghavendra--Tan~\cite{RaTa12} is somewhat weaker--looking at the \emph{mutual information} between the vectors rather than the orthogonal dot products. We only consider the stronger notion considered by Austrin--Benabbas--Georgiou~\cite{ABG16} in this paper.} if
\begin{equation}\label{eq:eps-uncor}
\quad \frac{1}{n^2}\sum_{i,j \in V}|\bv_i^{\perp} \cdot \bv_j^{\perp}| \LT \eps\;.
\end{equation}%
Note that \eqref{eq:eps-uncor} is not a traditional SDP constraint. Rather, one needs $O(1/\eps^4)$ rounds of the Sum of Squares (SoS) / Lasserre hierarchy to compute such a solution -- see Lemma 3.1 of \cite{ABG16} for the precise details. Given an $\eps$-uncorrelated solution, a much richer set of rounding schemes can be used and still ensure an almost balanced partition. In particular, given a set of \emph{rounding thresholds} $t_1, \hdots, t_n \in [-1, 1]$ such that $\sum_{i=1}^n t_i = 0$, we can perform the following rounding procedure:

\begin{itemize}
\item Sample $\br \in \mathcal N(0, I_{n+1})$. Let $\Phi : \mathbb R \to [0,1]$ be the CDF of the normal distribution with mean zero and variance one. For each $i \in V$, assign $i$ to $A$ if $\Phi(\bv_i^{\perp} \cdot \br) > \frac{1-t_i}{2}$ and to $B$ otherwise.
\end{itemize}

With high probability, we have that $(\frac12-\eps) n \le |A| \le (\frac12+\eps) n$ (see Lemma 3.6 of \cite{ABG16}). A small number of vertices can then be moved between $A$ and $B$ to get a bisection of the vertices without changing the cut value much. Much of the technical work in~\cite{ABG16} is spent on constructing elaborate procedures for choosing the thresholds $(t_1,t_2,\ldots,t_n)$ and analyzing their performance, with their best approximation ratio being $0.8776 \approx \agw - 10^{-3}$.

\subsection{Summary of Results}\label{sub-summary}

Austrin, Benabbas, and Georgiou \cite{ABG16} conjectured (henceforth, the ``ABG conjecture'') that the optimal approximate algorithm for MAX BISECTION should achieve essentially the same ratio as the Goemans-Williamson algorithm. More precisely, for every $\eps > 0$, there is an approximation algorithm for MAX BISECTION that achieves a ratio of $\agw - \eps$. The goal of this paper is to give evidence that the ABG conjecture might be false.

\begin{theorem}[Main Theorem, Informal]\label{thm:main-informal}
    Fix any $\eps > 0$. Then, for all sufficiently large $n$, there exists a graph $G = (V, E)$ on the vertex set $V = \{1, 2, \ldots, n\}$ and unit vectors $\bv_0, \bv_1, \ldots, \bv_n \in \mathbb{R}^{n+1}$  such that the following properties hold:
    \begin{enumerate}[(a)]
        \item $\sum_{i=1}^n \bv_0 \cdot \bv_i  = 0$.
        \item For $\{i,j\} \in E$, the triangle inequalities $-1 + |\bv_0 \cdot (\bv_i +\bv_j)| \le \bv_i \cdot \bv_j \le 1 - |\bv_0 \cdot (\bv_i-\bv_j)|$ hold.
        \item The vectors are $\eps$-uncorrelated, i.e., $\frac{1}{n^2}\sum_{i,j \in V}|\bv_i^{\perp} \cdot \bv_j^{\perp}| < \eps$.
        \item We have that $\OPT_{BIS}(G) \left/\left(\sum_{\{i,j\} \in E} \frac{1 - \bv_i \cdot \bv_j}{2}\right)\right. \le 0.87853 < \agw - 3 \cdot 10^{-5}$,
    \end{enumerate}
    where $\OPT_{BIS}(G)$ is the maximum size of a bisection of~$G$. %
\end{theorem}

To the best of our knowledge, this is also the first explicit example of a graph $G$ for which the qualitatively weaker bound $\OPT_{BIS}(G) / \SDP_{CUT}(G) < 0.87853$ holds. (Note that here we do not require the optimal solution of the Basic SDP relaxation of MAX BISECTION to be $\eps$-uncorrelated.) The claim also holds if the global constraint $\sum_{i,j=1}^n \bv_i\cdot \bv_j=0$, used by \cite{Ye01,HaZw02,FL06}, is added to the SDP relaxation of MAX CUT. %

We note that in the setting of MAX CSPs without a cardinality constraint, an integrality gap for a standard SDP (which does not have the $\eps$-uncorrelated condition) %
is sufficient to establish UGC hardness of the respective approximation problems~\cite{R08,austrin2010towards}. However, no analogous theorem is known for MAX CSPs with a global cardinality constraint. See the recent work of Ghoshal and Lee \cite{GhLe23} for a discussion of the state of the art for hardness results.

However, Raghavendra and Tan~\cite{RaTa12} gave a potential description of the algorithm/hardness boundary for approximate MAX CSPs with cardinality constraints in terms of \emph{dictatorship tests}. In particular, Raghavendra and Tan~\cite{RaTa12} show how integrality gaps like those of Theorem~\ref{thm:main-informal} can be converted into dictatorship tests. As a consequence, we have the following corollary.

\begin{corollary}[Dictatorship Test, informal]\label{cor:dict-test-informal}
There exists a dictatorship test for MAX BISECTION (in the sense of \cite{RaTa12}) whose ratio between soundness and completeness is less than $0.87853 < \agw - 3\cdot 10^{-5}$. 
\end{corollary}

\subsection{Related Results}

Austrin, Benabbas, and Georgiou \cite{ABG16}, improving on a result of Raghavendra and Tan \cite{RaTa12}, showed that MAX BISECTION 2-SAT, the balanced version of the MAX 2-SAT problem in which exactly half of the variables must be set to true, can be approximated to within a ratio of $\alpha_{LLZ}\simeq 0.9401$, which is the best ratio that can be obtained, under UGC, for the MAX 2-SAT problem. This ratio comes from an algorithm by Lewin, Livnat, and Zwick \cite{LLZ02} which was shown to be optimal for MAX 2-SAT under UGC by Austrin \cite{Austrin07,Austrin07_full}, assuming a certain simplicity conjecture. This simplicity conjecture was recently proved by Brakensiek, Huang, and Zwick \cite{BHZ24}. %

No loss in approximation ratio is incurred when moving from MAX 2-SAT to MAX BISECTION 2-SAT because the LLZ rounding scheme for MAX 2-SAT is of a special form that ensures balance when the SDP vectors are uncorrelated. Unfortunately, this `lucky coincidence' does not occur for MAX CUT and MAX BISECTION. %
In fact, Austrin and Stankovic \cite{AuSt19} showed that some MAX 2-CSP problems become harder when a global cardinality constraint is imposed. In particular, they showed that MAX CUT with the additional constraint that one side of the cut contains exactly $k$ vertices, for some $0<k<\frac{n}2$, can only be approximated to within a factor of $\alpha^{cc}_{cut}\simeq 0.858$, and that this is best possible under UGC. (Note that for MAX BISECTION we have $k=\frac{n}2$.) Similar results are also obtained for MAX 2-SAT with a global cardinality constraint. It is interesting to note that a balanced partition, i.e., $k=\frac{n}2$, is not the hardest case for both these problems. A special case of MAX 2-SAT with cardinality constraints is the MAX $k$-VC problem in which we are asked to find a subset of~$k$ vertices that covers as many edges as possible. For more details on MAX $k$-VC see \cite{AuSt19}.

MAX CSP problems without a global cardinality constraint are much better understood. Tight approximability results, assuming $P\ne NP$, are known for MAX 3-SAT and MAX 3-CSP \footnote{In the MAX 3-CSP problem, all constraints on at most three variables are allowed.} \cite{H01,KZ97,Zwick98,zwick02}. More generally, for all predicates which have a pairwise uniform distribution of solutions, it is NP-hard to obtain an approximation ratio which is better than taking a random guess \cite{SOC16}.

In the context of MAX CSPs, it is known by Raghavendra's theorem \cite{R08,R09} and the corresponding rounding algorithms due to Raghavendra and Steurer~\cite{RaSt09} that the optimal approximation ratio of \emph{any} MAX CSP is equal to the maximum possible integrality gap of the corresponding Basic SDP. However, these results do not tell us what the approximation ratio is for a given CSP or what the optimal rounding scheme looks like. For example, it is still an open problem to determine the optimal approximation ratios for problems such as MAX DI-CUT and MAX 2-AND, though it is known that under UGC, MAX DI-CUT cannot be approximated as well as MAX CUT \cite{BHPZ23}. (The goal of the line of research of the current paper is to obtain a similar separation between MAX CUT and MAX BISECTION.) Another major open problem in the area is to find the exact approximation ratio of MAX SAT, where clauses of all sizes are allowed--see the discussion in \cite{BHPZ25}. For additional results, see Charikar, Makarychev, and Makarychev \cite{CMM09} and the survey of Makarychev and Makarychev~\cite{MM17}.

Adapting the results of Raghavendra \cite{R08,R09} to cover MAX CSP problems with a global cardinality constraint is a challenging open problem. In particular, the results of \cite{RaTa12,ABG16} suggest that using the basic SDP relaxations for cardinality-constrained problems is not sufficient to obtain optimal approximation ratios. Much stronger SDP relaxations in the SoS/Lasserre hierarchy appear to be needed to obtain optimal, or close to optimal, approximation ratios for, e.g., the MAX BISECTION problem. For some results regarding globally cardinality constrained MAX CSP problems, and for further discussion, see Ghoshal and Lee \cite{GhLe23}.

Another major research avenue in recent years is to adapt the results of Raghavendra \cite{R08,R09} to \emph{satisfiable} instances of MAX CSP problems. Note that the Unique Games Conjecture is not a suitable assumption for hardness of approximation in the satisfiable regime, as satisfiable instances of the Unique Games problem can be easily recognized. Although there have been a number of attempts to come up with variants of the Unique Games Conjecture suitable for satisfiable instances (e.g., \cite{khot02,dinur2009conditional,brakensiek2021quest,Braverman21:itcs}), none have found the same universal applicability that UGC has had. Instead, recent works have sought to understand the boundary between dictatorship tests and algorithms for satisfiable CSPs. See for instance a number of works of Bhangale, Khot, Liu, and Minzer \cite{bhangale2023approximability,bhangale2024approximability,BKM25,bhangale2024approximability6,bhangale2024approximability7}. %

\subsection{Technical Overview}

To describe our integrality gap construction, we first recall the Basic SDP integrality gap obtained for MAX CUT. (We follow \cite{KhOd09,o2008optimal}. A slightly different derivation was given earlier in \cite{FeSc02}.). Recall that random hyperplane rounding performs the worst on pairs of vectors for which $\bv_i \cdot \bv_j = b_{GW} \approx -0.6891577$. For any natural number $N$, consider the continuous weighted graph whose vertex set is $\RR^N$. To construct the edges, sample pairs $\bx, \by\in \RR^N$ such that for all $i \in N$, $(x_i,y_i)$ is distributed according to a bivariate normal distribution with mean $(0,0)$ and covariance matrix $\begin{pmatrix}1 & b_{GW}\\b_{GW} & 1\end{pmatrix}$. The weight density of an edge $(\bx, \by)$ is precisely the probability density of $(\bx, \by)$ being drawn by the above process.

The analysis of this continuous integrality gap consists of two parts. First, for any $\eps > 0$, there is a sufficiently large~$N$, such that the instance has a Basic SDP solution $(\bv_\bx)_{\bx\in \RR^N}$ \footnote{Technically the Basic SDP is not defined for continuous instances, but this will be fixed by a discretization step.} with completeness approximately $\frac{1-b_{GW}}{2} - \eps$. This is done by setting $\bv_{\bx}=\frac{\bx}{\|\bx\|_2}$, for each $\bx \in \RR^N$. (Note that the unit vector $\frac{\bx}{\|\bx\|_2}$ is the normalization of~$\bx$.) Second, any cut of this continuous graph cuts at most $\frac{\cos^{-1}(b_{GW})}{\pi}$ edges. This is proved by applying a suitable Gaussian isoperimetric inequality~\cite{borell1985geometric} %
to show that the optimal cut must be a hyperplane cut through the origin, from which the bound of $\frac{\cos^{-1}(b_{GW})}{\pi}$ follows. Thus, for $N$ sufficiently large, the integrality gap of this continuous instance is at most $\agw + \eps$.

To convert this to a finite integrality gap, we \emph{discretize} the instance. In particular, we partition the unit sphere $\mathbb{S}^{N-1}$ (note that the Gaussian measure is concentrated near the unit sphere as $N \to \infty$) into approximately $(1/\eps)^N$ equal-volume pieces, such that each piece has diameter at most $\eps$. We then replace each piece by a single vertex. The weight between two regions is then the integral of the PDF across that pair of regions. With some careful bounding, it can be shown that the completeness and soundness of this finite discretized instance changes by at most $\eps$. Thus, we still have a $\agw + O(\eps)$ integrality gap which is finite. Note that while this instance is weighted, removing the weights on the edges can be done using standard techniques where we create a block of vertices for each original vertex and connect two blocks with a bipartite expander in which the number of edges is the same as the weight on the original edge~\cite{crescenzi2001weighted}. %

\paragraph{Blueprints.} Austrin~\cite{austrin2010towards} extended the above argument to obtain integrality gaps for more general MAX 2-CSP problems. (Some of the terms used below, including \emph{blueprint}, are coined by us here). A unit vector $\bv_0$ is used to represent false, i.e., the Boolean value~$0$. %
A triple of unit vectors $(\bv_0, \bv_1, \bv_2)$ can be summarized by the three dot products $b_1 := \bv_0 \cdot \bv_1, b_2 := \bv_0 \cdot \bv_2,$ and $b_{12} := \bv_1 \cdot \bv_2$. Since these vectors need to correspond to an assignment of Boolean variables, we impose \emph{triangle} inequalities, which can be summarized as $-1 + |b_1 + b_2| \leq b_{12} \leq 1 - |b_1 - b_2|$. We call such a triple $(b_1, b_2, b_{12}) \in [-1,1]^3$ a \emph{configuration}. We refer to $b_{12}$ as the \emph{pairwise bias} of the configuration.  For instance, the MAX CUT integrality gap just mentioned is based on the configuration $(0, 0, b_{GW})$. 

We define a \emph{blueprint} for a MAX 2-CSP problem to be a probability distribution of configurations. %
Note that each configuration $(b_1, b_2, b_{12})$ has a natural completeness value. For MAX CUT, this completeness is $\frac{1-b_{12}}{2}$. We can then define the completeness of the blueprint to be the weighted average of the completeness of the constituent configurations.

Defining the soundness of a blueprint is more subtle. For general blueprints, the soundness is determined by the best possible rounding scheme \cite{R08} which could be quite complicated. That said, a key result of Austrin~\cite{austrin2010towards} is that for blueprints for MAX 2-CSPs which are positive (for MAX CUT this means that ${b_1}{b_2} \geq b_{12}$, see \cite{austrin2010towards} for the general definition), it is sufficient to consider rounding schemes based on threshold functions.

More precisely, let $B$ be the set of biases appearing in a blueprint. A threshold function is a map $t : B \to [-1,1]$. For a given configuration $(b_1, b_2, b_{12})$ in the blueprint, we define the performance of $t$ on this configuration as follows. Let $\rho := \frac{b_{12} - b_1b_2}{\sqrt{1-b_1^2}\sqrt{1-b_2^2}}$ be the \emph{relative} pairwise bias of the configuration.\footnote{This assumes that $|b_1|,|b_2| \neq \pm 1$, which is true for all configurations considered in this paper.}  Let $\mathcal N_{\rho}$ be the bivariate normal distribution over $\RR^2$ with mean $(0,0)$ and covariance matrix $\begin{pmatrix}1&\rho\\\rho&1\end{pmatrix}$. We define the performance of $t$ on $(b_1,b_2,b_{12})$ to be
\[
    \Pr_{(g_1,g_2) \sim \mathcal N_{\rho}}\left[\one\left[g_1 > \Phi^{-1}\left(\frac{1-t(b_1)}{2}\right)\right] \neq \one\left[g_2 > \Phi^{-1}\left(\frac{1-t(b_2)}{2}\right)\right]\right].
\]
In other words, if we think of $t : B \to [-1,1]$ as a rounding scheme, we seek to know the probability that this scheme satisfies the CUT constraint. 
We define the performance of $t$ on a blueprint to be the average performance of $t$ on the constituent configurations. We define the \emph{soundness} of the blueprint to be the best performance of any threshold function $t : B \to [-1,1]$. We can thus define the \emph{approximation ratio} of the blueprint to be the ratio of the soundness to the completeness.

A key result of Austrin~\cite{austrin2010towards} is that for any Boolean MAX 2-CSP and for any blueprint which is positive and has an approximation ratio of $\alpha$, we can construct a finite integrality gap with an approximation ratio of $\alpha + \eps$ for any $\eps > 0$. In the case of MAX CSPs, integrality gaps are equivalent to UGC hardness~\cite{R08}, so this also yields (conditional) hardness of approximation.

\paragraph{The new blueprint for MAX BISECTION.} For MAX BISECTION (or other Boolean MAX 2-CSPs with a cardinality constraint), we can adapt Austrin's technique by extending the notion of a blueprint. 
To capture the bisection constraint, a blueprint should also include a probability distribution $\mu$ over the set of biases $B$ used by the blueprint such that the mean of biases under~$\mu$ is zero. When computing the soundness of such a blueprint, we require that any threshold function $t : B \to [-1,1]$ which is used has mean zero under~$\mu$. We do this to ensure that the proportion of vertices on each side of the cut will be equal. %
A key technical contribution in this paper is the construction of a blueprint for MAX BISECTION whose approximation ratio is strictly less than $\agw$. For simplicity in this overview, we describe the blueprint to 3 digits of precision. A more precise description of the blueprint is given in Table~\ref{table:blueprint} in Section~\ref{sub-blueprint}.

The set of biases $B$ is $\{-0.635, -0.315, 0.004, 0.307, 0.626\}$. Our probability distribution $\mu$ over~$B$ samples $-0.635$ with probability $0.326$ and $0.307$ with probability $0.674$. All other biases are sampled with probability zero. The distribution of configurations in the blueprint is as follows.
\begin{center}
\begin{tabular}{cr}
\multicolumn{1}{c}{probability} & \multicolumn{1}{c}{bias}\\
\hline
$0.326$ & $-0.635$  \\
$0$ & $-0.315$ \\
$0$ & $0.004$  \\
$0.674$ & $0.307$ \\
$0$  & $0.626$  
\end{tabular}
\hspace*{15pt}
\begin{tabular}{c@{\quad\quad(\;}r@{\;,\;}r@{\;,\;}c@{\;\;)}}
\multicolumn{1}{c}{probability\quad\quad} & \multicolumn{3}{c}{configuration}\\
\hline
0.351 & $0.307$ & $-0.315$ & $b_{GW}$ \\
0.274 & $0.004$ & $0.307$ & $b_{GW}$ \\
0.272 & $0.004$ & $-0.315$ & $b_{GW}$ \\
0.065 & $0.626$ & $-0.315$ & $b_{GW}$ \\
0.039 & $-0.635$ & $0.626$ & $b_{GW}$
\end{tabular}
\end{center}
The soundness to completeness ratio of this blueprint is about $0.878523\simeq \agw-4.4\times 10^{-5}$. This is only slightly below $\agw$ but this is enough to show that an approximation ratio of $\agw$ for MAX~BISECTION cannot be obtained by just using the paradigm of \cite{RaTa12} and \cite{ABG16}. We also note that biases with zero probability are still useful, and in fact play a key role in the construction. That said, in the actual reduction we may slightly perturb the blueprint so that every bias has nonzero weight, without affecting the approximation ratio much (see Corollary~\ref{cor:blueprint_mu_robust} for details).

We remark that this blueprint was discovered in part by adapting the minimax algorithm of \cite{BHPZ23} for algorithmically discovering blueprints for MAX 2-CSPs.

\paragraph{Intuition for the blueprint.}
A key reason that this blueprint gives an integrality gap instance is that once we choose the threshold for bias $-0.635$, the balance constraint fixes the threshold for bias $0.307$ and makes it larger than it would be if we used hyperplane rounding. The $(0.004,0.307,b_{GW})$ configurations penalize this larger threshold as for these configurations, it would be better to ignore the biases and use hyperplane rounding on the components of the vectors which are orthogonal to~$\bv_0$.

To see why the threshold for bias $0.307$ is larger than it would be under hyperplane rounding, we make the following observations:
\begin{enumerate}
\item When $x$ is small, $\Phi(x) \approx \frac{1}{2} + \frac{1}{\sqrt{2\pi}}(x - \frac{x^3}{6})$ as 
\[
\frac{1}{\sqrt{2\pi}}\int_{-\infty}^{x}{e^{-\frac{x^2}{2}}dx} = 
\frac{1}{2} + \frac{1}{\sqrt{2\pi}}\int_{0}^{x}{e^{-\frac{x^2}{2}}dx} \approx \frac{1}{2} + \frac{1}{\sqrt{2\pi}}\int_{0}^{x}{1-\frac{x^2}{2}dx} = \frac{1}{2} + \frac{1}{\sqrt{2\pi}}(x - \frac{x^3}{6})
\]
This implies that when $t = \frac{1}{2} + r$ and $r$ is small, $\Phi^{-1}(t) \approx \sqrt{2\pi}r + \frac{(\sqrt{2\pi})^{3}}{6}r^3$ which is close to being linear in $r$.
\item If we use hyperplane rounding then for a vector $\bv = b\bv_0 + \sqrt{1 - b^2}\bv^{\perp}$, if we sample a random vector $\bu$ and we have that $\bu \cdot \bv_0 = z$ then the threshold for $\bv^{\perp}$ will be $x = -\frac{bz}{\sqrt{1 - b^2}}$. This means that if we have two vectors $\bv_1$ and~$\bv_2$ where $\bv_1 = {b_1}\bv_0 + \sqrt{1 - b_1^2}\bv_1^{\perp}$ and $\bv_2 = {b_2}\bv_0 + \sqrt{1 - b_2^2}\bv_2^{\perp}$ then if we sample a random vector $\bu$ and we have that $\bu \cdot \bv_0 = z$, the thresholds for $\bv_1^{\perp}$ and~$\bv_2^{\perp}$ will be $x_1 = -\frac{{b_1}z}{\sqrt{1 - {b_1}^2}}$ and $x_2 = -\frac{{b_2}z}{\sqrt{1 - {b_2}^2}}$. Note that $\frac{x_1}{x_2} = \frac{b_1}{b_2} \cdot \frac{\sqrt{1 - b_2^2}}{\sqrt{1 - b_1^2}} \approx (1 - \frac{b_2^2 - b_1^2}{2})\frac{b_1}{b_2}$ so compared to a rounding scheme where $r$ is linear in $b$, hyperplane rounding gives thresholds with slightly smaller magnitudes for vectors with smaller biases.
\end{enumerate}

In Appendix~\ref{app:theory}, we give some more insight into why there is a gap between MAX BISECTION and MAX CUT by giving a rough theoretical analysis showing that there is a gap between MAX BISECTION and MAX CUT when we allow negated variables. Note that this is a weaker statement than our main result as having negated variables forces us to have that $t(-b) = -t(b)$ for all $b \in [-1,1]$. Also note that since this theoretical analysis requires the biases to have small magnitude, it gives a much smaller gap.

\paragraph{Analyzing the blueprint.}

The completeness of our blueprint is precisely $\frac{1 - b_{GW}}{2}$ since every configuration in the blueprint has a pairwise bias equal to $b_{GW}$. The soundness is much more difficult to analyze. We claim the following

\begin{lemma}\label{lem:sound-blueprint}
There exists $\delta > 10^{-5}$ such that the soundness of the blueprint is less than $\frac{\cos^{-1}(b_{GW})}{\pi} - \delta$. In other words, for every $t : B \to [-1,1]$ such that $\sum_{b \in B} \mu(b)t(b) = 0$, the performance of $t$ is at most $\frac{\cos^{-1}(b_{GW})}{\pi} - \delta$.
\end{lemma}

In other words, the approximation ratio of our MAX BISECTION blueprint is \emph{strictly} less than $\agw$. To verify Lemma~\ref{lem:sound-blueprint}, we perform a computer-assisted proof using \emph{interval arithmetic}. In particular, we adapt the interval arithmetic routines used in recent works on approximation algorithms for MAX CSPs \cite{BHPZ23,BHZ24} which are built on the Arb interval arithmetic library in FLINT \cite{johansson2017arb,johansson2018numerical,johansson2019computing}. See Appendix~\ref{appendix:ia} for further details. Note that although the approximation ratio of our MAX BISECTION blueprint is only slightly below $\agw$, we obtain a rigorous computer-assisted proof that it is indeed strictly smaller than $\agw$, proving the main claim of this paper.

\paragraph{Conversion into an Integrality Gap instance.}

To convert the blueprint into an integrality gap instance (and thus prove Theorem~\ref{thm:main-informal}), we perform a reduction similar to that of Austrin~\cite{austrin2010towards} and O'Donnell-Wu~\cite{o2008optimal}.%

In particular, we consider a sufficiently large natural number $N$ and construct a continuous instance whose vertex set is $V = B \times \RR^N$. To construct the weighted edges, we look at every configuration $(b_1, b_2, b_{12})$ in the blueprint. Recall that the relative pairwise bias of such a configuration is $\rho = \frac{b_{12}-b_1b_2}{\sqrt{(1-b_1^2)(1-b_2^2)}}$. We sample pairs $\bx_1, \bx_2 \in \RR^N$ such that for each $i \in [N]$, $(x_{1,i}, x_{2,i})$ is an independent sample from $\mathcal N_{\rho}$. For such pairs $(\bx_1, \bx_2)$, we connect $(b_1, \bx_1)$ and $(b_2, \bx_2)$ with an edge whose weight is proportional to the probability density of $(\bx_1, \bx_2)$ times the probability of $(b_1, b_2, b_{12})$ in the blueprint. For the (weighted) bisection constraint, we also given each $(b, \bx) \in V$ a weight proportional to $\mu(b)$.

Similar to the completeness and soundness analysis for MAX CUT, we show for any $\eps > 0$ there is some $N$ such that the completeness and soundness of this continuous instance equals those of the blueprint up to $\pm \eps$ (see Lemma~\ref{lem:gaussian_mixture} for more details). %
As such, for $\eps \ll 10^{-5}$, our continuous instance is an integrality gap for weighted MAX BISECTION with ratio strictly less than $\agw$.

To complete the proof, we still need to (1) convert the continuous instance into a suitable finite instance, and (2) make all the vertices and edges unweighted. We achieve (1) as in the case of MAX~CUT by discretizing each unit sphere $\{b\} \times \mathbb{S}^{N-1} \subseteq B \times \mathbb{S}^{N-1}$ (note that the Gaussian measure will be concentrated near the unit sphere) into a finite number of regions, with the corresponding edge weight between regions to be the integral of the weights in continuous instance over the pair of regions. The vertex weight of a region that lies in $\{b\} \times \mathbb{S}^{N-1}$ will be $\mu(b)$, the probability mass on~$b$ in the original blueprint. Here, we may apply some small perturbation to $\mu$ so that every vertex gets a weight bounded away from 0 (see Corollary~\ref{cor:blueprint_mu_robust} for details). %

Given this finite instance with weighted vertices and edges, we convert it into an unweighted graph by adapting a standard construction due to Creszenzi, Silvestri, and Trevisan~\cite{crescenzi2001weighted}, %
where we replace each vertex of the finite graph with a ``cloud'' of vertices, and each weighted edge with the unweighted edges of a bipartite expander graph. Here the size of the ``cloud'' is proportional to the weight of the original vertex, and the number of edges in the expander graph is proportional to the weight of the original edge. It is straightforward to show that the SDP value of this instance can only increase, while the soundness is approximately preserved via the expander mixing lemma. As such, the approximation ratio is (roughly) the same, so we still have an integrality gap with ratio strictly less than $\agw$. %

Note that Corollary~\ref{cor:dict-test-informal} follows immediately from this integrality gap, as Raghavendra and Tan~\cite{RaTa12} showed that any integrality gap for MAX BISECTION can be converted into a suitable dictatorship test. See Corollary~\ref{cor:dict_test} for the precise formulation.

\subsection*{Outline}

In Section~\ref{sec:prelim}, we give the necessary background on Gaussian distributions, SDP relaxations (including the Sum of Squares hierarchy) and blueprints. In Section~\ref{sec:bisection}, we present the exact blueprint we use for MAX BISECTION, verify its approximation ratio, and work out the corresponding integrality gap. In Section~\ref{sec:conclusion}, we summarize our results and list some open questions.

In Appendix~\ref{appendix:ia}, we describe the interval arithmetic verification  in detail. %
In Appendix~\ref{app:theory}, we give a rough theoretical analysis showing that there is a gap between MAX BISECTION and MAX CUT when we allow negated variables. %

\section{Preliminaries}\label{sec:prelim}

Let $\agw = \min_{b \in [-1,1)} \frac{\cos^{-1}(b)}{\pi} \cdot \frac{2}{1 - b} \approx 0.8785672$ be the approximation ratio achieved by Goemans-Williamson. Let $\bgw = \arg\min_{b \in [-1,1)} \frac{\cos^{-1}(b)}{\pi} \cdot \frac{2}{1 - b} \approx -0.6891577$ be the \emph{critical} bias in which the minimum is obtained, and let $\cgw=\frac{1-\bgw}{2}\approx 0.844578$ be the completeness at the critical bias.

\subsection{Facts about Gaussian and spherical distributions}

Let $\Phi$ and $\varphi$ be the c.d.f.\ and p.d.f.\ for the standard Gaussian distribution. Let $\mathcal{N}(0, I_d)$ denote the standard $d$-dimensional Gaussian distribution. For any $\rho \in [-1, 1]$, we use $\bx, \by \sim_\rho \mathcal{N}(0, I_d)$ to denote the following sampling procedure: for every $i \in [d]$, independently sample $(x_i, y_i) \sim \mathcal{N}\left(\left(\begin{array}{cc}
     0  \\
     0 
\end{array}\right), \left(\begin{array}{cc}
     1 & \rho  \\
     \rho & 1
\end{array}\right)\right)$, return $\bx = (x_1, \ldots, x_d), \by = (y_1, \ldots, y_d)$. 

For any $q_1, q_2 \in [0, 1]$, we let
\[ \Gamma_\rho(q_1, q_2) = \Pr_{(x,y) \sim_\rho N(0, 1)}[x \leq \Phi^{-1}(q_1) \text{ and } y \leq \Phi^{-1}(q_2)]\;.\]

We record some partial derivatives for $\Gamma_\rho(q_1, q_2)$ here for later use. The derivatives with respect to~$q_1$ and~$q_2$ can be easily derived from the definitions and the derivative with respect to $\rho$ can be found in Drezner and Wesolowsky \cite{DW90}.

\begin{lemma}\label{lem:Gamma_partials}
    We have the following derivatives for $\Gamma_\rho(q_1, q_2)$:
\begin{align*}
\frac{\partial}{\partial q_1} \Gamma_{\rho}(q_1, q_2) &\EQ \Phi\left(\frac{\Phi^{-1}(q_2)  - \rho \Phi^{-1}(q_1)}{\sqrt{1-\rho^2}}\right),\\
\frac{\partial}{\partial q_2} \Gamma_{\rho}(q_1, q_2) &\EQ \Phi\left(\frac{\Phi^{-1}(q_1)  - \rho \Phi^{-1}(q_2)}{\sqrt{1-\rho^2}}\right),\\
\frac{\partial}{\partial \rho} \Gamma_{\rho}(q_1, q_2) &\EQ \frac{1}{2\pi \sqrt{1-\rho^2}} \exp\left(-\frac{\Phi^{-1}(q_1)^2 - 2\rho \Phi^{-1}(q_1)\Phi^{-1}(q_2) + \Phi^{-1}(q_2)^2}{2(1-\rho^2)}\right).
\end{align*}
\end{lemma}

We will need the following useful lemma, whose proof can be found in, e.g.,~\cite{Vershynin_2018}.
\begin{lemma}\label{lem:inner_preserving}
    Let $\rho \in [-1, 1]$ and $\bx, \by \sim_\rho \mathcal{N}(0, I_d)$, then 
    \[
    \Pr[|\bx \cdot \by - \rho| \geq \eps] \LE 2 \exp(-\Omega(\eps^2 d))\;.
    \]
\end{lemma}

We state a powerful theorem on Gaussian isoperimetry due to Borell~\cite{borell1985geometric}. The statement here is adapted from~\cite{austrin2010towards}. 

\begin{theorem}[\cite{borell1985geometric}]\label{thm:borell}
    Let $\rho \in [-1, 1]$. Let $f, g: \mathbb{R}^n \to [0, 1]$ be two integrable functions with $\E_{\bx \sim \mathcal{N}(0, I_n)}[f(\bx)] = q_1$ and $\E_{\by \sim \mathcal{N}(0, I_n)}[g(\by)] = q_2$. Then $\E_{\bx, \by \sim_\rho\, \mathcal{N}(0, I_n)}[f(\bx)g(\by)] \leq \Gamma_\rho(q_1, q_2)$ if $\rho \geq 0$, and $\E_{\bx, \by \sim_\rho\, \mathcal{N}(0, I_n)}[f(\bx)g(\by)] \geq \Gamma_\rho(q_1, q_2)$ if $\rho \leq 0$.
\end{theorem}

We will also need the following fact about the uniform distribution over the unit sphere $\mathbb{S}^{d-1}=\{\bx\in\RR^d \mid \|\bx\|_2=1\}$. The proof can be found in~\cite{Vershynin_2018}.

\begin{theorem}
    Let $\bu\in\mathbb{S}^{d-1}$ and $\delta > 0$. %
    Then for a uniformly sampled $\bv \in \mathbb{S}^{d-1}$, we have
    \begin{equation}
        \Pr_\bv[\bu \cdot \bv \geq \delta] \LE 2\exp\left(-\delta^2d/2\right).
    \end{equation}
\end{theorem}

\begin{corollary}\label{cor:uncorrelated}
    Let $\bu\in\mathbb{S}^{d-1}$. Then for a uniformly sampled $\bv \in \mathbb{S}^{d-1}$, we have
    \begin{equation}
        \E_\bv\bigl[|\bu \cdot \bv|\bigr] \EQ O\left(\sqrt{\frac{\log d}{d}}\right).
    \end{equation}
\end{corollary}

\subsection{SDP relaxations and rounding schemes}

As we discussed in Section~\ref{sec:rt_abg}, the paradigm for the state-of-the-art MAX BISECTION algorithm starts by solving the \emph{$\ell$-round Sum-of-Squares/Lasserre SDP hierarchy}. Given a graph $G = (V, E)$, where $V=\{1,2,\ldots,n\}$, the $\ell$-round Sum-of-Squares/Lasserre SDP can be formulated as follows (this formulation is taken from~\cite{ABG16}):
\begin{alignat}{3}
&\text{maximize} &\qquad& \sum_{\{i, j\} \in E}  \frac{1 - \bv_i\cdot\bv_j}{2} \notag \\ 
&\text{subject to} &\qquad& \bv_0 \cdot \sum_{i \in V}\bv_{S \oplus \{i\}} = 0\;, &\qquad&\textstyle \forall\, S\in \binom{V}{<\ell} \\
& &\qquad&\bv_{S_1} \cdot \bv_{S_2} = \bv_{S_3} \cdot \bv_{S_4}\;,&\qquad&\textstyle\forall\, S_1,S_2,S_3,S_4 \in \binom{V}{\le\ell}\; ,\; S_1 \oplus S_2 = S_3 \oplus S_4 \;, \\
& &\qquad& \bv_S \cdot \bv_S = 1\;,&\qquad&\textstyle\forall\,  S\in \binom{V}{\le\ell}\;.
\end{alignat}
Here $\binom{V}{<\ell}$ and $\binom{V}{\le\ell}$ denote the set of all subsets of~$V$ of size less than~$\ell$, or at most~$\ell$, respectively, and $S\oplus T$ denotes the \emph{symmetric difference} of $S$ and $T$. In this SDP we have a unit vector $\bv_S$ for every set $S \in \binom{V}{\le\ell}$, and it is intended to simulate the scalar $\prod_{i \in S}x_i$ where $\bx\in \{-1,+1\}^V$ represents some bisection of $G$. We use $\bv_0$ as an alias for $\bv_\varnothing$, and $\bv_i$ as a shorthand for $\bv_{\{i\}}$.

\begin{definition}[Cf.\ Definition 2.2 in~\cite{ABG16}]
    We say that unit vectors $\bv_0, \bv_1, \ldots, \bv_n$ are an $\eps$-uncorrelated MAX BISECTION vector solution for a graph $G = (V, E)$, where $V=[n]$, if 
    \begin{equation}
        \sum_{i = 1}^n \bv_0 \cdot \bv_i \EQ 0 \quad\text{and}\quad\frac{1}{n^2}\sum_{i, j \in V}\left|\bv_i^\perp \cdot \bv_j^\perp\right| \LE \eps\;,
    \end{equation}
    where $\bv_i^{\perp} = \frac{\bv_i - (\bv_0 \cdot \bv_i) \bv_i}{\sqrt{1 - (\bv_0 \cdot \bv_i)^2}}$. (We assume that $|\bv_0\cdot \bv_i|\ne 1$.)
    Furthermore, we say that $\bv_0, \bv_1, \ldots, \bv_n$ satisfy the triangle inequalities on the edges if for every $\{i, j\} \in E$, we have
    \begin{equation}
        -1 + |\bv_0 \cdot \bv_i + \bv_0 \cdot \bv_j| \LE \bv_i \cdot \bv_j \LE 1 - |\bv_0 \cdot \bv_i - \bv_0 \cdot \bv_j|\;.
    \end{equation}
\end{definition}

\begin{remark}\label{remark:weighted_uncorrelated}
     The above definition can be easily generalized to a vertex-weighted version by requiring $\sum_{i = 1}^n w_i\, \bv_0 \cdot \bv_i = 0$ and $\frac{1}{(\sum_{i \in V} w_i)^2}\sum_{i, j \in V}w_iw_j\left|\bv_i^\perp \cdot \bv_j^\perp\right| \leq \eps$, where $w_i$ is the weight on vertex $i$.
\end{remark}

A $\THRESH^-$ rounding scheme~\cite{LLZ02} is specified by a continuous function $t:[-1,1] \to [-1, 1]$. The rounding scheme chooses a standard random Gaussian vector $\br \sim \mathcal{N}(0, I_{n+1})$, and sets variable~ $i$ to $1$ if $\br \cdot \bv_i^\perp \geq \Phi^{-1}\left(\frac{1 - t(\bv_0 \cdot \bv_i)}{2}\right)$ and $-1$ otherwise. For any vertices $i$ and $j$, the probability that $i$ and $j$ get different signs from this rounding scheme is then equal to
\begin{align}
    & \Pr\left[\br \cdot \bv_i^\perp \geq \Phi^{-1}\left(\frac{1 - t(\bv_0 \cdot \bv_i)}{2}\right) \text{ and } \br \cdot \bv_j^\perp \leq \Phi^{-1}\left(\frac{1 - t(\bv_0 \cdot \bv_j)}{2}\right)\right] \;\;+ \\
     &  \Pr\left[\br \cdot \bv_i^\perp \leq \Phi^{-1}\left(\frac{1 - t(\bv_0 \cdot \bv_i)}{2}\right) \text{ and } \br \cdot \bv_j^\perp \geq \Phi^{-1}\left(\frac{1 - t(\bv_0 \cdot \bv_j)}{2}\right)\right] \phantom{\;\;+} \\
    =\;\;& \frac{1-\bv_0 \cdot \bv_i}{2} + \frac{1-\bv_0 \cdot \bv_j}{2} - 2\Gamma_{\rho}\left(\frac{1-t(\bv_0 \cdot \bv_i)}{2}, \frac{1-t(\bv_0 \cdot \bv_j)}{2}\right),\label{eq:thresh_sound}
\end{align}
where we used the fact that $\br \cdot \bv_i^\perp, \br \cdot \bv_j^\perp \sim_\rho \mathcal{N}(0, 1)$ with $\rho = \bv_i^\perp \cdot \bv_j^\perp$. 

We may also take a distribution of $\THRESH^-$ rounding schemes. The class of rounding algorithms where we take such a distribution was named $\THRESH$ by~\cite{LLZ02}.%

\subsection{Configurations and blueprints for hard MAX BISECTION instances}

\begin{definition}
    A configuration for MAX BISECTION is a triple $(b_i, b_j, b_{ij}) \in [-1, 1]^3$ satisfying the triangle inequality 
    \begin{equation}
        -1 + |b_i + b_j| \LE b_{ij} \LE 1 - |b_i - b_j| \;.
    \end{equation}
\end{definition}

A configuration $(b_i, b_i, b_{ij})$ represents the three inner products $(\bv_0\cdot \bv_i,\bv_0\cdot \bv_j,\bv_i\cdot \bv_j)$, where $\bv_0,\bv_i,\bv_j$ are three unit vectors appearing in the vector solution of an SDP relaxaion of MAX BISECTION that imposes the triangle inequalities. We refer to $b_i$ and $b_j$ as \emph{biases}, and $b_{ij}$ as the \emph{pairwise bias}.

\begin{definition}
    Let $\theta = (b_i, b_j, b_{ij})$ be a configuration. Its \emph{relative pairwise bias} is defined to be $\rho(\theta) \coloneqq \frac{b_{ij} - b_ib_j}{\sqrt{(1-b_i^2)(1 - b_j^2)}}$ if $(1-b_i^2)(1 - b_j^2) \neq 0$, and $0$ otherwise.
\end{definition}

The geometric meaning of the relative pairwise bias is that $\rho = \bv_i^\perp \cdot \bv_j^\perp$. Austrin introduced the following notion of \emph{positive configurations}. Intuitively, for these configurations the optimal rounding scheme must come from $\THRESH^-$. 

\begin{definition}[cf.\ Definition 3.4 in~\cite{austrin2010towards}]
    A configuration $\theta = (b_i, b_j, b_{ij}) \in [-1, 1]^3$ for MAX BISECTION is said to be \emph{positive} if $\rho(\theta) \leq 0$.
\end{definition}

We introduce the notion of \emph{blueprints} for hard MAX BISECTION instances. Blueprints generalize the idea of distributions of configurations used by~\cite{austrin2010towards} for proving UG-hardness of MAX 2-CSPs without cardinality constraints. Similar notions are also used in \cite{BHPZ23}.

\begin{definition}
    A \emph{blueprint} is a triple $\mathcal{D} = (D, B, \mu)$, where $D$ is a finite distribution of configurations, $B$ is the set of biases that appear in $D$, and $\mu$ is a probability measure over $B$ such that $\sum_{b \in B} \mu(b)\,b  = 0$. We say that the blueprint is \emph{positive} if every configuration in the support of $D$ is positive.
\end{definition}
We now define completeness and soundness of blueprints.

\begin{definition}[Completeness of blueprints] The completeness of a blueprint $\mathcal{D} = (D, B, \mu)$ is defined as follows:
\[ \comp(\mathcal{D}) \EQ \E_{(b_i,b_j,b_{ij})\sim D} \frac{1-b_{ij}}{2}. \]
\end{definition}

Intuitively, if the distribution of configurations in the blueprint comes from an actual SDP solution, then the completeness is equal to the SDP value achieved by the solution. To define soundness of blueprints, we first define the notion of balance for blueprints.

\begin{definition}[$\eps$-almost balanced]
    A threshold function $t: B \to [-1, 1]$ is \emph{$\eps$-almost balanced} for a blueprint $\mathcal{D} = (D, B, \mu)$ if $\left|\sum_{b \in B} \mu(b)t(b)\right| \leq \eps $. We say $t$ is balanced for $\mathcal{D}$ if $\sum_{b \in B} \mu(b)t(b) = 0$.
\end{definition}

We call $t$ a threshold function since it should be thought of as a function that specifies a $\THRESH^-$ rounding scheme for the blueprint. If the blueprint comes from an actual SDP solution, then the $\THRESH^-$ rounding scheme specified by an $\eps$-almost balanced threshold function will produce a bisection (up to an $\eps$ fraction) with high probability.

\begin{definition}[Soundness of blueprints]
    Given a blueprint $\mathcal{D} = (D, B, \mu)$ and a threshold function $t: B \to [-1, 1]$, we define \begin{equation}
        \sound(\mathcal{D}, t) \EQ \E_{\theta=(b_i,b_j,b_{ij})\sim D} \left[\frac{1-t(b_i)}{2} + \frac{1-t(b_j)}{2} - 2\Gamma_{\rho(\theta)}\left(\frac{1-t(b_i)}{2}, \frac{1-t(b_j)}{2}\right)\right]\;.
    \end{equation}
    Here the expression inside the expectation is taken from~\eqref{eq:thresh_sound}. We also define \begin{equation}
        \sound(\mathcal{D}) \EQ \max_{t:\, t \text{ balanced for } \mathcal{D}}\sound(\mathcal{D}, t)\;,
    \end{equation}
    and
    \begin{equation}
        \sound_\eps(\mathcal{D}) \EQ \max_{t:\, t\,\, \eps\text{-almost balanced for } \mathcal{D}}\sound(\mathcal{D}, t)\;.
    \end{equation}
    Note that here both maxima are well-defined by compactness.
\end{definition}

We now prove some simple but useful facts about the soundness. Essentially, these facts say that the soundness is robust against small perturbations.

\begin{lemma}\label{lem:soundness_lipshitz}
    Let $\mathcal{D} = (D, B, \mu)$ be a blueprint. For any two threshold functions $t_1, t_2: B \to [-1, 1]$, we have
    \begin{equation}
        |\sound(\mathcal{D}, t_1) - \sound(\mathcal{D}, t_2)| \LE \sum_{b \in B} |t_1(b) - t_2(b)|\;.
    \end{equation}
\end{lemma}
\begin{proof}
    Using Lemma~\ref{lem:Gamma_partials} it can be easily verified that $|\frac{\partial}{\partial t(b)} \sound(\mathcal{D}, t)| \leq 1$ for any $t$ and $b$.
\end{proof}

\begin{corollary}\label{cor:blueprint_eps_robust}
     Let $\mathcal{D} = (D, B, \mu)$ be a blueprint. For any $\eps > 0$, we have 
     \begin{equation}         \sound(\mathcal{D})\LE\sound_\eps(\mathcal{D}) \LE \sound(\mathcal{D}) + |B|\eps\;.
     \end{equation}
\end{corollary}
\begin{proof}
    The first inequality follows directly from definition. To show the second inequality, let $t: B \to [-1, 1]$ be an $\eps$-almost balanced threshold function for $\mathcal{D}$ such that $\sound_\eps(\mathcal{D}) = \sound(\mathcal{D}, t)$. Note that we can always find another threshold function $t'$ that is balanced for $\mathcal{D}$ which satisfies $|t'(b) - t(b)| \leq \eps$ for every $b \in B$. It then follows from Lemma~\ref{lem:soundness_lipshitz} that
    \begin{align*}
        \sound_\eps(\mathcal{D}) \EQ \sound(\mathcal{D}, t) & \LE \sound(\mathcal{D}, t') + |B| \eps\\
        & \LE \sound(\mathcal{D}) + |B| \eps\;.\qedhere
    \end{align*}
\end{proof}

\begin{corollary}\label{cor:blueprint_mu_robust}
     Let $\mathcal{D} = (D, B, \mu)$ be a blueprint. Assume that $B$ contains both positive and negative biases, then for any sufficiently small $\eps > 0$, there is another blueprint $\mathcal{D}' = (D, B, \mu')$ such that
     \begin{itemize}
         \item $\comp(\mathcal{D}) = \comp(\mathcal{D}')$,
         \item $\sound(\mathcal{D'})\leq \sound(\mathcal{D}) + O(\eps)$,
         \item For every $b \in B$, $\mu'(b) \geq \eps$.
     \end{itemize}
     Here $O(\cdot)$ hides a constant which depends only on $B$.
\end{corollary}
\begin{proof}
    By our assumption that $B$ contains both positive and negative biases, for every $b \in B$, we may find another distribution $\mu_{b}$ on $B$ such that $\mu_{b}(b) \geq c_b$, and $\sum_{b' \in B}\mu_{b}(b') = 0$, where $c_b > 0$ is some constant depending only on $B$. We
    define 
    \[
    \mu'(b) \EQ (1 - \eps)\mu(b) + \frac{\eps}{|B|}\left(\sum_{b' \in B}\mu_{b'}(b)\right)\;,
    \]
    and let $\mathcal{D}' = (D, B, \mu')$. Then for every $b$, $\mu'(b) \geq c_b\eps/|B|$. Since completeness is defined independently from $\mu$, we have $\comp(\mathcal{D}) = \comp(\mathcal{D}')$. For the soundness, note that if $t$ is balanced for $\mathcal{D'}$, then we have 
    \[
    \sum_{b \in B} t(b)\left((1 - \eps)\mu(b) + \frac{\eps}{|B|}\left(\sum_{b' \in B}\mu_{b'}(b)\right)\right) \EQ 0\;,
    \]so
    \begin{equation}
        \left|\sum_{b \in B} t(b)\mu(b)\right| \EQ \frac{\eps }{(1 - \eps)|B|} \left|\sum_{b \in B}t(b)\left(\sum_{b' \in B}\mu_{b'}(b)\right)\right|\LE 2\eps|B|.
    \end{equation}
    By Corollary~\ref{cor:blueprint_eps_robust}, we then have
    $\sound(\mathcal{D'})\leq \sound_{2\eps|B|}(\mathcal{D})\leq \sound(\mathcal{D}) + 2\eps|B|^2$. The claim then follows by rescaling $\eps$ appropriately so that $\mu'(b) \geq \eps$ for every $b \in B$.
\end{proof}

\section{BISECTION appears to be harder than CUT}\label{sec:bisection}

In this section, we prove our main theorem.

\begin{theorem}\label{thm:main}
    Fix any $\eps > 0$. Then for all sufficiently large $n$, there exists a graph $G = (V, E)$ on the vertex set $V = \{1, 2, \ldots, n\}$ satisfying the following properties:
    \begin{enumerate}[(a)]
        \item There is an $\eps$-uncorrelated MAX BISECTION vector solution $\bv_0, \bv_1, \ldots, \bv_n$ for $G$ satisfying the triangle inequalities on the edges such that $\sum_{\{i, j\} \in E} \frac{1 - \bv_i \cdot \bv_j}{2} \geq (\cgw - O(\eps)) |E|$,
        \item For every $A \subseteq V$ such that $(\frac12 - \eps)n \leq |A| \leq (\frac12 + \eps)n$, $|\{ \{i, j\} \in E \mid i \in A, j \in V \setminus A\}| \leq (0.87853 \cdot \cgw + O(\eps)) |E|$.
    \end{enumerate}
\end{theorem}

Theorem~\ref{thm:main} gives strong evidence that approximating MAX BISECTION might be harder than approximating MAX CUT. At the very least, it shows that the current algorithmic paradigm of rounding SDP vectors with low correlation cannot achieve the same ratio as Goemans-Williamson. In particular, the approximation ratio that we can achieve this way is at most $0.87853 < \alpha_{GW}$. We note that in our construction, the SDP vectors only satisfy the triangle inequalities when there is an edge present. We argue that this is not be a big issue, due to the following reasons:
\begin{itemize}
    \item For MAX CUT, it is known that integrality gap instances for the SDP without triangle inequalities can be converted into integrality gap instances for the SDP with triangle inequalities via Unique Games hardness results~\cite{KhVi05}.
    \item The current best algorithm due to~\cite{ABG16} does not use triangle inequalities when no edge is present.
    \item The dictatorship test constructed by Raghavendra and Tan~\cite{RaTa12} only requires that there is a local distribution on every edge (the existence of local distribution over an edge $\{i, j\}$ is equivalent to the satisfaction of triangle inequalities involving $\bv_0, \bv_i, \bv_j$), which our instance satisfies. We state the properties of this dictatorship test as Corollary~\ref{cor:dict_test}. For the exact formulation of this dictatorship test, we refer interested readers to~\cite{RaTa12}.
\end{itemize}

\begin{corollary}\label{cor:dict_test}
    There exist absolute constants $C, K$ such that the following holds. For every $\eta, \tau \in (0, 1)$ and integer $R > 0$, there exists a dictatorship test $\mathsf{DICT}$ for functions $f: \{-1, 1\}^R \to \{-1, 1\}$ such that 
    \begin{itemize}
        \item If $f$ is a dictator function, then $\mathsf{DICT}(f)$ accepts with probability at least $\cgw - 2\eta$.
        \item If $f$ has influence at most $\tau$ on every coordinate, then $\mathsf{DICT}(f)$ accepts with probability at most $0.87853 \cdot \cgw + C \cdot \tau^{K \eta}$.
    \end{itemize}
\end{corollary}
\begin{proof}
The corollary follows directly from Theorem~\ref{thm:main}, and Theorem 6.4 in~\cite{RaTa12}.
\end{proof}

The remainder of this section will be devoted to the proof of Theorem~\ref{thm:main}. To this end, we will first construct a positive blueprint for MAX BISECTION with a large completeness versus soundness gap. Intuitively, this means that the blueprint is hard to round with $\THRESH^-$ rounding schemes that respect the bisection condition. The blueprint was found by a computer search using a minimax method similar to the one used in~\cite{BHPZ23}. 

We then present a general reduction that converts positive blueprints into integrality gap instances. Using ideas from~\cite{o2008optimal}, we first convert a blueprint into an integrality gap instance where the graph is continuous (and weighted). We then obtain a finite (and weighted) integrality gap instance by taking an $\eps$-net, following~\cite{FeSc02} and~\cite{RaSt09}. Finally, we obtain an unweighted integrality gap instance following a reduction of~\cite{crescenzi2001weighted}. The main technical difference in our reduction is that we need to respect the bisection condition in all these steps.

\subsection{A MAX BISECTION blueprint hard for $\THRESH$}\label{sub-blueprint}

\begin{lemma}\label{lem:hard_dist}
    There exists a positive blueprint $\mathcal{D}^* = (D, B, \mu)$ such that:
    \begin{itemize}
        \item $\comp(\mathcal{D}^*) = \cgw$,
        \item (Verified using interval arithmetic) $\sound(\mathcal{D}^*) < 0.87853 \cdot \cgw$.
    \end{itemize}
    It follows that $\alpha(\mathcal{D}^*)=\sound(\mathcal{D}^*)/\comp(\mathcal{D}^*) < 0.87853$.
\end{lemma}

\begin{proof}
The description of the blueprint can be found in Table~\ref{table:blueprint}, and a graph representation for it can be found in Figure~\ref{fig:enter-label}. Note that in the second configuration we have $-b_3-b_4+\bgw=-1$, in the third we have $b_2+b_3+\bgw=-1$, and in the forth $-b_2-b_5+\bgw=-1$. These are all tight triangle inequalities. (The parameters $\nu_1 = 0.004$ and $\nu_2 = 0.013$ can probably be optimized a bit, but the current choices are close to being optimal and the effect of such an optimization is expected to be tiny. The choice of~$b=1+\bgw$ is believed to be optimal for this type of blueprints.)

\begin{table}[H]
\centering
\begin{tabular}{c@{\quad\quad}c}
\begin{tabular}{cc@{\hspace{4pt}}c@{\hspace{4pt}}cr}
\multicolumn{1}{c}{$\mu$} & \multicolumn{3}{c}{bias} \\
\hline
$0.325898600625$ & $b_1$  $=$ & $-2b-\nu_2$  & $\approx$ $-0.634684524710$ \\
$0$ & $b_2$  $=$ & $-b-\nu_1$   & $\approx$ $-0.314842263355$ \\
$0$ & $b_3$  $=$ & $\nu_1$      & $=$       $\phantom{+}0.004000000000$ \\
$0.674101399375$ & $b_4$  $=$ & $b - \nu_1$  & $\approx$ $\phantom{+}0.306842263355$ \\
$0$ & $b_5$  $=$ & $2b + \nu_1$ & $\approx$ $\phantom{+}0.625684524710$ \\
\end{tabular}
&
\hspace*{-10pt}
\begin{tabular}{c@{\quad\quad(\;}c@{\;,\;}c@{\;,\;}c@{)}}
\multicolumn{1}{c}{probability\quad\quad\quad} & \multicolumn{3}{c}{configuration}\\
\hline
0.351359472465 & $b_2$ & $b_4$ & $b_{GW}$ \\
0.273707303709 & $b_3$ & $b_4$ & $b_{GW}$ \\
0.271584315668 & $b_2$ & $b_3$ & $b_{GW}$ \\
0.064406822738 & $b_2$ & $b_5$ & $b_{GW}$ \\
0.038942085420 & $b_1$ & $b_5$ & $b_{GW}$
\end{tabular}
\end{tabular}
\[ b = 1 + b_{GW} \approx 0.310842263355 \quad,\quad \nu_1 = 0.004 \quad,\quad \nu_2 = 0.013  \]
\vspace*{-20pt}
\caption{The MAX BISECTION blueprint $\mathcal{D}^*$ with $\alpha(\mathcal{D}^*)< 0.87853$. (Non-rigorous numerical calculations suggest that $\alpha(\mathcal{D}^*)\approx 0.878523 \approx \agw - 4.4\times 10^{-5}$.)
}
\label{table:blueprint}
\end{table}

We have $\comp(\mathcal{D}^*) = \cgw$ since every $\theta = (b_i, b_j, b_{ij})$ in the support of $D$ satisfies $b_{ij} = \bgw$. The soundness claim is formally verified using interval arithmetic. We explain the high-level ideas for the verification here, leaving the details to Appendix~\ref{appendix:ia}.  

By definition, we need to verify that for any choice of $t: B \to [-1, 1]$ that is balanced for $\mathcal{\mathcal{D}^*}$, the soundness $\sound(\mathcal{D}^*, t)$ is less than $0.87853 \cdot \cgw$. Observe that if we fix $t(b_1)$, then the value of $t(b_4)$ is uniquely determined by the balance condition since $b_1$ and $b_4$ are the only two biases with nonzero weights in $\mu$. We can therefore write $\sound(\mathcal{D}, t)$ as a $4$-variate function $s(t_1, t_2, t_3, t_5)$ where $t_i = t(b_i) \in [-1, 1]$ ($t_4$ is determined by $t_1$ via the bisection constraint). Verifying the maximum of a $4$-variate function to 5 decimal places can still be very time-consuming, so we further reduce it to a 2-parameter problem as follows.

We claim that if in addition to $t_1$ we also fix $t_2$, then there exists a unique $t_3$ that maximizes the soundness. This can be seen by computing the partial derivative with respect to $t_3$:
\begin{align*}
    \frac{\partial s}{\partial t_3} \EQ\,\, & w(b_3, b_4) \cdot \left(-\frac{1}{2} + \Phi\left(\frac{\Phi^{-1}\left(\frac{1 - t_4}{2}\right) - \rho(b_3, b_4)\Phi^{-1}\left(\frac{1 - t_3}{2}\right)}{\sqrt{1 - \rho(b_3, b_4)^2}}\right)\right) \\
     & \quad +  w(b_2, b_3) \cdot \left(-\frac{1}{2} + \Phi\left(\frac{\Phi^{-1}\left(\frac{1 - t_2}{2}\right) - \rho(b_2, b_3)\Phi^{-1}\left(\frac{1 - t_3}{2}\right)}{\sqrt{1 - \rho(b_2, b_3)^2}}\right)\right).
\end{align*}
Here $w(b_i, b_j)$ denotes the probability weight on the configuration $(b_i, b_j, b_{GW})$ in $D$, and $\rho(b_i, b_j) = \rho(b_i, b_j, \bgw)$. Note that since $\rho(b_i, b_j) < 0$ for every configuration in $D$, $\frac{\partial s}{\partial t_3}$ is monotonically decreasing in $t_3$. This means that there is a unique value in $[-1, 1]$ for $t_3$ that maximizes $s$ (ignoring the cases where $t_2, t_4 = \pm 1$, which can be dealt with easily). This value can be found by a variant of the binary search procedure that tries to find a root for $\frac{\partial s}{\partial t_3}$. We discuss the details of the implementation of this procedure in Appendix~\ref{appendix:ia}.

In a similar way, we can find the unique optimal choice for $t_5$. In summary, by fixing some arbitrary~$t_1$ and~$t_2$, we can algorithmically find the unique optimal choices for the remaining three parameters. The problem then becomes certifying an upper bound for the maximum of a bivariate function $s(t_1, t_2)$. We provide a contour plot of this function in Figure~\ref{fig:blueprint_soundness_contour}. Non-rigorous numerical experiments suggest that this function is at most $0.8785231...$, and we verified $s(t_1, t_2) / \cgw < 0.87853$ rigorously using interval arithmetic.
\end{proof}

\begin{figure}\label{fig:blueprint}
    \centering
    
\begin{tikzpicture}
    \def\numvertices{5}
    \def\radius{2.5cm} %
    \def\nodesize{0.8cm} %

    \foreach \label/\vertexname in {$b_3$/1, $b_2$/2, $b_1$/3, $b_5$/4, $b_4$/5} {
        \node (\vertexname) at ({360/\numvertices * (\vertexname - 1) + 90}:\radius) [
            circle,
            draw=black,      %
            fill=white,      %
            text=black,      %
            minimum size=\nodesize
        ] {\label}; %
    }

    \draw[thick] (1) -- (5);
    \draw[thick] (1) -- (2);
    \draw[thick] (2) -- (5);
    \draw[thick] (3) -- (4);
    \draw[thick] (2) -- (4);
    \draw[dotted, thick] (3) -- (5);

\end{tikzpicture}
    \caption{A graph representation for the blueprint $\mathcal{D}^*$. A solid line between two biases means there is a configuration with these two biases. The dotted line connects the two biases with nonzero bias weights.}
    \label{fig:enter-label}
\end{figure}
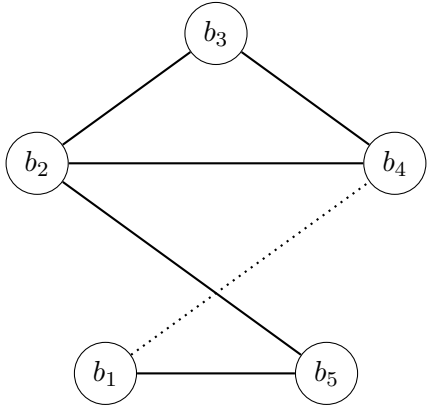

\begin{figure}[h]
    \centering
    \includegraphics[scale=0.6]{contour.png}
    \vspace*{-20pt}
    \caption{A contour plot for $s(t_1, t_2) / \cgw$. Non-rigorous numerical experiments suggest that this function is at most $0.8785231...$, and we verified $s(t_1, t_2) / \cgw < 0.87853$ rigorously using interval arithmetic. %
    }
    \label{fig:blueprint_soundness_contour}
\end{figure}

Note that in $\mathcal{D}^*$, the support of $\mu$ is a strict subset of $B$. Furthermore, equality is attained in some of the triangle inequalities. To make the blueprint more amenable to our reduction, we can add some small perturbations as in the following lemma.

\begin{lemma}\label{lem:blueprint_delta_perturb}
    Let $\mathcal{D}^* = (D, B, \mu)$ be as in Lemma~\ref{lem:hard_dist}. For every sufficiently small $\delta > 0$, there exists a positive blueprint $\mathcal{D}^*_\delta = (D', B, \mu')$ such that the following holds:
    \begin{itemize}
        \item $\comp(\mathcal{D}^*_\delta) \geq\comp(\mathcal{D}^*) - O(\delta)$.
        \item $\sound(\mathcal{D}^*_\delta) \leq \sound(\mathcal{D}^*) + O(\delta)$.
        \item For every $(b_i, b_j, b_{ij})$ in the support of D', we have that the triangle inequalities are strict, i.e., $-1 + |b_i + b_j| < b_{ij} < 1 - |b_i - b_j|$.
        \item For every $b \in B$, $\mu'(b) \geq \delta$.
    \end{itemize} 
\end{lemma}
\begin{proof}
    Since $D$ is finitely supported, we may add $\delta > 0$ to all pairwise biases (namely, replace $b_{ij}$ by $b_{ij} + \delta$) while still satisfying the triangle inequalities as long as $\delta$ is sufficiently small. 
    This changes the completeness value by $O(\delta)$, and furthermore, since $\rho(\theta)$ is strictly bounded away from $-1$ for every $\theta \in D$, we have by Lemma~\ref{lem:Gamma_partials} that the soundness changes by up to $O(\delta)$. We can then use Corollary~\ref{cor:blueprint_mu_robust} to perturb $\mu$ and get $\mu'$ such that $\mu'(b) \geq \delta$ for every $b \in B$. This again only changes the soundness by $O(\delta)$.
\end{proof}

\subsection{Gaussian mixture graph}

Given a blueprint $\mathcal{D} = (D, B, \mu)$, we construct an continuous \emph{Gaussian mixture graph} $\mathcal{G}_\mathcal{D}^d$, building upon the ideas of~\cite{KhOd09,o2008optimal}. The vertex set of the Gaussian mixture graph is $B \times \mathbb{R}^d$. The edge weights are generated using the following sampling procedure: first sample a configuration $\theta = (b_i, b_j, b_{ij}) \sim D$, then sample a pair of $\rho(\theta)$-correlated $d$-dimensional Gaussian vectors $\bv_i^\perp, \bv_j^\perp\in \RR^d$, and return $((b_i, \bv_i^\perp), (b_j, \bv_j^\perp))$. The completeness of $\mathcal{G}_\mathcal{D}^d$ is defined to be 
\begin{equation}
\comp(\mathcal{G}_\mathcal{D}^d) \coloneqq \sup_{f: B \times \mathbb{R}^d \to \mathbb{S}^{d-1}} \E_{((b_i, \bv_i^\perp), (b_j, \bv_j^\perp))\sim \mathcal{G}_\mathcal{D}^d}\frac{1 - f(b_i, \bv_i^\perp)\cdot f(b_j, \bv_j^\perp)}{2}.
\end{equation}

We say that an assignment $f: B \times \mathbb{R}^d \to [-1, 1]$ for $\mathcal{G}_\mathcal{D}^d$ is \emph{$\eps$-almost balanced} for $\mathcal{D}$ if 
\[
\left|\sum_{b \in B} \mu(b)\E_{\bu \sim N(0, I_d)}[f(b,\bu)]\right| \LE \eps.
\]
The \emph{$\eps$-robust soundness} of $\mathcal{G}_D^d$ is defined to be 
\begin{equation}
\sound_\eps(\mathcal{G}_\mathcal{D}^d) \coloneqq \sup_{\substack{f: B \times \mathbb{R}^d \to [-1, 1] \\ f \text{ is }\eps\text{-almost balanced}}} \E_{((b_i, \bv_i), (b_j, \bv_j))\sim \mathcal{G}_\mathcal{D}^d}\frac{1 - f(b_i, \bv_i)f(b_j, \bv_j)}{2}.
\end{equation}

\begin{lemma}\label{lem:gaussian_mixture}
    Let $\mathcal{D} = (D, B, \mu)$ be a positive blueprint and $d > 0$ a positive integer. Then we have 
    \begin{equation}
        \comp(\mathcal{G}_\mathcal{D}^d) \GE \comp(\mathcal{D}) - o_d(1),
    \end{equation}
    and
    \begin{equation}
        \sound_\eps(\mathcal{G}_\mathcal{D}^d) \LE \sound_\eps(\mathcal{D}).
    \end{equation}
\end{lemma}
\begin{proof}
    For the completeness, consider the mapping $g: \bx \mapsto \frac{\bx}{\|\bx\|}$ ($\|\bx\|$ denotes the Euclidean norm of~$\bx$). It is a standard probability exercise (see, e.g., Theorem 4.3 in the full version of~\cite{o2008optimal}) to show that 
    \begin{equation}
        \left|\E_{(\bu, \bv) \sim \mathcal{G}_{\mathcal{D}}^d}\left[\bu \cdot \bv - g(\bu)\cdot g(\bv)\right]\right| \LE O\left(\sqrt{\frac{\log d}{d}}\right).
    \end{equation}
    As for the soundness, let $f: B \times \mathbb{R}^d \to [-1, 1]$ be any assignment that is $\eps$-almost balanced for~$\mathcal{D}$. Consider the threshold function defined by $t: b \mapsto \E_{\bu \sim N(0, I_d)}[f(b, \bu)]$. Then $t$ is also $\eps$-almost balanced for $\mathcal{D}$. We have
    \begin{align*}
        &\,\,\E_{((\bv_i)_{b_i}, (\bv_j)_{b_j})\sim \mathcal{G}_\mathcal{D}^d}\frac{1 - f(b_i, \bv_i)f(b_j, \bv_j)}{2}\\
        \EQ&\,\, \E_{((\bv_i)_{b_i}, (\bv_j)_{b_j})\sim \mathcal{G}_\mathcal{D}^d}\left[\frac{1 - f(b_i, \bv_i)}{2} + \frac{1 - f(b_i, \bv_i)}{2} - 2 \cdot \frac{(1 - f(b_i, \bv_i))}{2}\cdot\frac{(1 - f(b_j, \bv_j))}{2}\right] \\
        \LE&\,\, \frac{1 - t(b_i)}{2} + \frac{1 - t(b_j)}{2} - 2\cdot\Gamma_{\rho(\theta)}\left(\frac{1 - t(b_i)}{2}, \frac{1 - t(b_j)}{2}\right) \\
        \EQ&\,\, \sound(\mathcal{D}, t).
    \end{align*}
    Here the inequality follows from Theorem~\ref{thm:borell} and the fact that $\rho(\theta) \leq 0$ for every $\theta$ in the support of $D$. The soundness claim then follows by taking the supremum over all $\eps$-almost balanced $f$.
\end{proof}

\subsection{Discretization via $\eps$-nets}

We now use \emph{$\eps$-nets} to discretize the continuous instance. This is a standard technique in the construction of integrality gap instances and rounding algorithms for MAX CSPs without cardinality constraints (see e.g., ~\cite{FeSc02,o2008optimal,RaSt09}). The main difference in our proof is that we need to maintain the bisection condition for soundness. We use the following lemma regarding $\eps$-nets.

\begin{lemma}[See, e.g., Lemma 21 in~\cite{FeSc02}]\label{lem:eps_net}
    For every sufficiently small $\eps > 0$, the sphere $\mathbb{S}^{d-1}$ can be partitioned into $m = m(d, \eps) = (O(1) / \eps)^d$ equal-volume cells $C_1, \ldots, C_m$ such that for every $i \in [m]$ and every $\bu, \bv \in C_i$, $\|\bu - \bv\|_2 \leq \eps$.
\end{lemma}

\begin{theorem}\label{thm:eps_net_reduction}
    Let $\mathcal{D} = (D, B, \mu)$ be a positive blueprint. Assume that for every configuration $(b_i, b_j, b_{ij})$ in the support of $D$, the triangle inequalities are strict, i.e., $-1 + |b_i + b_j| < b_{ij} < 1 + |b_i - b_j|$. Fix a sufficiently small $\eps > 0$. Then, for every sufficiently large $n$, there exists a vertex-weighted, edge-weighted graph $G = (V, E, w_V, w_E)$ on the vertex set $V = \{1, 2, \ldots, n\}$ such that the following holds:
    \begin{itemize}
        \item There exists an $O(\eps)$-uncorrelated MAX BISECTION vector solution (in the weighted sense as in Remark~\ref{remark:weighted_uncorrelated}) $\bv_0, \ldots, \bv_n$ for $G$ satisfying triangle inequalities on the edges such that 
        \begin{equation}
            \sum_{\{i, j\} \in E} w_E(\{i, j\})  \frac{1 - \bv_i \cdot \bv_j}{2} \GE \comp(\mathcal{D}) - O(\eps).
        \end{equation}
        \item For every $A \subseteq V$ such that $|\sum_{i \in A} w_V(i) - \sum_{i \in V \setminus A} w_V(i)| \le \delta \cdot(\sum_{i \in V}w_V(i))$, we have that 
        \begin{equation}
            \sum_{i \in A, j \in V \setminus A} w_E(\{i, j\}) \LE \sound(\mathcal{D}) + O(\delta) + O(\eps).
        \end{equation}
    \end{itemize}
    Here $O(\cdot)$ hides some constant independent of $\eps$ and $\delta$.
\end{theorem}
\begin{proof}
Let $d > 0$ be some large integer whose value we will choose later. By Lemma~\ref{lem:eps_net}, we can divide $\mathbb{S}^{d-1}$ into $m = m(d, \eps)$ equal volume cells $\mathcal{C} = \{C_1, \ldots, C_m\}$ so that any two vectors within the same cell have distance at most $\eps$. Let $f : \mathbb{S}^{d-1} \to \mathcal{C}$ be the function which given any $\bx \in \mathbb{S}^{d-1}$, returns the cell $C \in \mathcal{C}$ that contains $\bx$.
    We take the vertex set $V$ to be $(B \times \mathcal{C}) \cup \{v, v'\}$\footnote{We may rename the vertices using $1, \ldots, n$ later.} where~$v$ and~$v'$ are two extra auxiliary vertices. For the vertex weights, for a vertex $(b, C) \in (B \times \mathcal{C})$ we set $w_V((b, C)) = \mu(b)$, and for $v, v'$ we set $w_V(v) = w_V(v') = 1$. The edges and edge weights in $G$ are defined via the following sampling procedure: 
    \begin{itemize}
        \item Sample $\theta = (b_1, b_2, b_{12}) \sim D$.
        \item Sample $\bx, \by \sim_{\rho(\theta)} \mathcal{N}(0, I_d)$. 
        \item If any of the following three conditions, $|\|\bx\|^2 - 1| \geq \eps$, $|\|\by\|^2 - 1| \geq \eps$, $|\bx \cdot \by - \rho(\theta)| \geq \eps$, are met, then return $\{v, v'\}$.
        \item Otherwise, return $\{\,(b_1, f(\frac{\bx}{\|\bx\|}))\,,\, (b_2, f(\frac{\by}{\|\by\|}))\,\}$.
    \end{itemize}
    The edge set $E$ is the set of edges which the above procedure samples with nonzero probability, and we use these probabilities as their edge weights $w$. 

    To define the vector solution, we first choose some arbitrary unit vector $\bv_0$ (which will be orthogonal to all vectors we choose later). For every $C \in \mathcal{C}$, we pick an arbitrary unit vector $\bv_C^\perp \in C$, and we define $\bv_{(b,C)} \coloneqq b\bv_0 + \sqrt{1 - b^2} \bv_C^\perp$ to be the SDP vector for the vertex $(b,C)$. For the auxiliary variables $v$ and $v'$, we simply choose some unit vectors that are orthogonal to every other vector in the vector solution. 
    We now verify that this is indeed an $O(\eps)$-uncorrelated vector solution for $G$ in the weighted sense (we ignore $v, v'$ since they are negligible compared to the rest of the graph). First, we have
    \begin{equation}
        \sum_{(b, C)\in (B \times \mathcal{C})} w_V((b, C)) \bv_{(b, C)} \cdot \bv_0 \EQ \sum_{(b, C)\in (B \times \mathcal{C})} \mu(b)b \EQ |\mathcal{C}| \cdot \sum_{b \in B}\mu(b)b \EQ 0 \;.
    \end{equation}
    Now, for every $C \in \mathcal{C}$, we have 
    \begin{align*}
        \frac{1}{|\mathcal{C}|}\sum_{C' \in \mathcal{C}} |\bv_C^\perp \cdot \bv_{C'}^\perp| & \LE \int_{\bu \in \mathbb{S}^{d-1}}|\bv_C^\perp \cdot \bu| + |\bv_C^\perp \cdot (\bv_{C'}^\perp-\bu)| \mathrm{d}\bu \\
        & \LE O\left(\sqrt{\frac{\log d}{d}}\right) + \eps\;.
    \end{align*}
    Here the last inequality follows from Corollary~\ref{cor:uncorrelated} and the definition of an $\eps$-net. By choosing $d$ sufficiently large, we can make the above expression at most $2\eps$, thereby establishing the $O(\eps)$-uncorrelated property of vectors.

    For a pair of random vectors $\bx, \by \sim_{\rho} \mathcal{N}(0, I_d)$, we say that $\bx$ and $\by$ are $\eps$-good if $|\|\bx\|^2 - 1| < \eps$, $|\|\by\|^2 - 1| < \eps$, and $|\bx \cdot \by - \rho| < \eps$. Note that by Lemma~\ref{lem:inner_preserving}, for any $\eps > 0$, we have $\bx$ and $\by$ are $\eps$-good with probability $1 - \eps$ if $d$ is sufficiently large. For any $\theta = (b_i, b_j, b_{ij}) \in D$, and two random vectors $\bx, \by \sim_{\rho(\theta)} \mathcal{N}(0, I_d)$, conditioning on the event that $\bx$ and $\by$ are $\eps$-good, we have that
    \begin{equation}
        \left|\bv_{f(\frac{\bx}{\|\bx\|})}^\perp\cdot \bv_{f(\frac{\by}{\|\by\|})}^\perp - \rho(\theta)\right| \EQ O(\eps)\;,
    \end{equation}
    from which it follows that
    \begin{equation}
        \left|\bv_{(b_i, f(\frac{\bx}{\|\bx\|}))} \cdot \bv_{(b_j, f(\frac{\by}{\|\by\|}))} - b_{ij}\right| \EQ O(\eps)\;.
    \end{equation}
    
So we have that completeness and the triangle inequalities are preserved up to $O(\eps)$. Since the triangle inequalities are strict for every $\theta \in D$, we can preserve triangle inequalities for every edge by making $\eps$ sufficiently small.
    
    For the soundness, assume that $A \subseteq V$ is such that 
    \begin{equation}
    \left|\sum_{i \in A} w_V(i) - \sum_{i \in V \setminus A} w_V(i)\right| \LE \delta \cdot\left(\sum_{i \in V}w_V(i)\right) \EQ \delta \cdot (|\mathcal{C}| + 2)\;,   
    \end{equation}
    where we have the +2 term to account for the auxiliary vertices $v$ and $v'$.  Consider the assignment $h_A: B \times \mathbb{R}^d \to [-1, 1]$ for the Gaussian mixture graph $\mathcal{G}_{\mathcal{D}}^d$ where $h_A(b, \bx) = 1$ if $(b, f(\frac{\bx}{\|\bx\|})) \in A$ and~$-1$ otherwise. We first verify that $h_A$ is almost balanced. We have
    \begin{equation}\label{eq:h_estimate}
    \sum_{b \in B} \mu(b)\E_{\bu \sim \mathcal{N}(0, I_d)}[h_A(b, \bu)] \EQ \sum_{b \in B} \mu(b) \cdot \frac{|(\{b\}\times \mathcal{C}) \cap A| - |(\{b\}\times \mathcal{C}) \setminus A|}{|\mathcal{C}|}
    \end{equation}
    
    By the condition on the size of $A$ we have
    \begin{equation}\label{eq:A_estimate}
        \left|\sum_{b \in B} \mu(b) \cdot\left( |(\{b\}\times \mathcal{C}) \cap A| - |(\{b\}\times \mathcal{C}) \setminus A|\right)\right| \LE \delta(|\mathcal{C}| + 2) + 2\;,
    \end{equation}
    Combining \eqref{eq:A_estimate} and \eqref{eq:h_estimate}, we obtain
    \begin{equation}
        \left| \sum_{b \in B} \mu(b)\E_{\bu \sim \mathcal{N}(0, I_d)}[h_A(b, \bu)]\right| \LE \frac{\delta(|\mathcal{C}| + 2) + 2}{|\mathcal{C}|} \leq \delta + \eps\;.
    \end{equation}
    Here the last inequality holds as long as $d$ (and therefore $|\mathcal{C}|$) is sufficiently large. So $h_A$ is $(\delta + \eps)$-almost balanced for $\mathcal{G}_{\mathcal{D}}^d$. By Corollary~\ref{cor:blueprint_eps_robust} and Lemma~\ref{lem:gaussian_mixture}, we have
    \begin{align*}
    & \E_{\theta=(b_1,b_2,b_{12}) \sim D}\;\E_{\bx,\by \sim_{\rho(\theta)}\, \mathcal{N}(0, I_d)} \left[\frac{1 - h(b_1,\bx)h(b_2,\by)}{2}\right] \\
    \LE\,\,& \sound_{\delta + \eps}(\mathcal{G}_\mathcal{D}^d) \leq \sound_{\delta + \eps}(\mathcal{D}) \leq \sound(\mathcal{D}) + (\delta + \eps)|B|
    \end{align*}
    
    It follows that
    \begin{align*}
        & \,\quad \sum_{i \in A, j \in V \setminus A} w_E(\{i, j\}) \\
        & \LE 
        \E_{\theta=(b_1,b_2,b_{12}) \sim D}\;\E_{\bx,\by \sim_{\rho(\theta)}\, \mathcal{N}(0, I_d)} \left[\frac{1 - h(b_1,\bx)h(b_2,\by)}{2} \cdot \bone[\bx, \by \text{ is }\eps\text{-good}]+ \bone[\bx, \by \text{ is not }\eps\text{-good}]\right] \\ 
        & \LE \E_{\theta=(b_1,b_2,b_{12}) \sim D}\;\E_{\bx,\by \sim_{\rho(\theta)}\, \mathcal{N}(0, I_d)} \left[\frac{1 - h(b_1,\bx)h(b_2,\by)}{2}\right] + \eps\\
        & \LE \sound(\mathcal{D}) + \delta|B| + \eps(|B|+1)\;.
    \end{align*}
    This completes the proof.
\end{proof}

A weighted version of our main theorem follows directly from the reduction in Theorem~\ref{thm:eps_net_reduction}.
\begin{theorem}\label{thm:main_thm_weighted}
    Fix any sufficiently small $\eps, \eta > 0$. Then, for any sufficiently large $n$, there exists a vertex-weighted, edge-weighted graph $G = (V, E, w_V, w_E)$ on the vertex set $V = \{1, 2, \ldots, n\}$ such that the following properties hold:
    \begin{enumerate}[(a)]
        \item There is an $O(\eps)$-uncorrelated MAX BISECTION vector solution $\bv_0, \bv_1, \ldots, \bv_n$ for $G$ satisfying the triangle inequalities on the edges such that 
        \begin{equation}
        \sum_{\{i, j\} \in E} w_E(\{i,j\})\cdot\frac{1 - \bv_i \cdot \bv_j}{2} \GE (\cgw - O(\eps) - O(\eta)) \sum_{\{i, j\} \in E} w_E(\{i,j\})\;,
        \end{equation}
        \item For every $A \subseteq V$ such that $(\frac12 - \eps)(\sum_{i \in V}w_V(i)) \LE \sum_{i \in A}w_V(i) \leq (\frac12 + \eps)(\sum_{i \in V}w_V(i))$, we have
        \begin{equation}
            \sum_{\{i, j\} \in E, i \in A, j \in V \setminus A} w_E(\{i,j\}) \LE (0.87853 \cdot \cgw + O(\eps) + O(\eta))\sum_{\{i, j\} \in E} w_E(\{i,j\})\;.
        \end{equation}
        \item For every $i \in V$, $w_V(i) \in [\eta, 1]$.
    \end{enumerate}
    Here $O(\cdot)$ hides some constant that is independent of $\eps$ and $\eta$.
\end{theorem}
\begin{proof}
    This follows by applying the reduction in Theorem~\ref{thm:eps_net_reduction} to the blueprint from Lemma~\ref{lem:blueprint_delta_perturb}.
\end{proof}

\subsection{Weighted vs. unweighted MAX BISECTION}
We complete the proof of our main theorem by converting the vertex-weighted, edge-weighted graph into an unweighted graph. To this end, we adapt a reduction by Crescenzi, Silvestri, and Trevisan~\cite{crescenzi2001weighted}, which they used for MAX CSPs without cardinality constraints. Here we crucially use the fact that the vertex weights we obtained in Theorem~\ref{thm:main_thm_weighted} are bounded away from 0.

\begin{definition}[cf.\ Definition 12 in~\cite{crescenzi2001weighted}]
Let $w \in \mathbb{N}^+$ and $\eps, \delta>0$. We say that a bipartite (unweighted) graph $G = (V_1, V_2, E)$ is $(m, n, w, \delta, \eps)$-mixing if $|V_1| = m$, $|V_2|=n$, and for any $A \subseteq V_1$, $B \subseteq V_2$, if $|A| \geq \delta|V_1|$, $|B| \geq \delta|V_2|$, then 
\[
\left||E(A, B)| - w \frac{|A||B|}{mn}\right| \LE \eps w \frac{|A||B|}{mn}\;,
\]
where $E(A, B)$ is the set of edges in $E$ with one endpoint in $A$ and another in $B$.
\end{definition}

Using explicit construction of expanders, we can obtain the following lemma.

\begin{lemma}[cf.\ Theorem 13 in~\cite{crescenzi2001weighted}]\label{lem:expander_mixing}
    Fix some $\eta > 0$. For every $\eps, \delta > 0$, there exist $c, n_0$ such that for every $m, n > n_0$, $\min(m/n, n/m) \geq \eta$, and every $w \in [c\min(m,n), mn]$, there is an $(m, n, w, \delta, \eps)$-mixing bipartite graph that can be constructed in time $poly(m,n)$.
\end{lemma}

We now describe a version of the CST reduction which we adapted to handle the cardinality constraints. The reduction demonstrates the equivalence between unweighted MAX BISECTION and vertex-weighted, edge-weighted MAX BISECTION in terms of their approximability, assuming that the vertex weights are bounded between $[t, 1]$ for some $t > 0$.

\begin{theorem}[cf.\ Theorem 14 in~\cite{crescenzi2001weighted}]\label{thm:weighted_unweighted_reduction}
    Fix some $\eta > 0$. Assume that there is an $\alpha$-approximation algorithm for unweighted MAX BISECTION problem. Then for every $\tau > 0$, there exists an $(\alpha+\tau)$-approximation algorithm for the vertex-weighted, edge-weighted MAX BISECTION problem\footnote{In this vertex-weighted version, we only require the partition $(A, V \setminus A)$ to satisfy $|\sum_{i \in A} w_V(i) - \sum_{i \not\in A} w_V(i)| \leq 1$. Note that finding an exact bisection with respect to the weights is at least as hard as the subset sum problem.}, where the vertex weights are bounded between $[\eta, 1]$ and edge weights are positive integers bounded by some polynomial in the number of vertices.
\end{theorem}
\begin{proof}
Let $\eps, \delta > 0$ be some constants depending on $\eta$ that we will choose later. Let $c, n_0$ be the constants guaranteed by Lemma~\ref{lem:expander_mixing}. Let $G = (V, E, w_V, w_E)$ be a weighted graph where $w_V: V \to [\eta, 1]$ gives the vertex weights and $w_E: E \to \mathbb{Z}^+$ gives the edge weights. As in~\cite{crescenzi2001weighted}, we may assume that there exists some $N \geq n_0$ polynomial in $|V|$ such that $c\eta N \leq w_E(e) \leq N^2$ for every $e \in E$.

We construct a new unweighted graph $G' = (V', E')$ as follows:
\begin{itemize}
    \item The vertex set is $V' = \bigcup_{i \in V}\left( \{i\} \times [\lceil w_V(i) N \rceil]\right)$. In words, for each vertex $i \in V$ in the original graph, we create $\lceil w_V(i) N \rceil$ vertices for it in $G'$. 
    \item For each $\{i, j\} \in E$, we construct a copy of $(\lceil w_V(i) N \rceil, \lceil w_V(j) N \rceil, w_E(\{i, j\}), \delta, \eps)$-mixing graph and place it between $\{i\} \times [\lceil w_V(i) N \rceil]$ and $\{j\} \times [\lceil w_V(j) N \rceil]$.
\end{itemize} 

Note that the size of $G'$ is polynomial in the size of $G$. Observe that any vertex set $A \subset V$ in $G$ that satisfies $|\sum_{i \in A} w_V(i) - \sum_{i \not\in A} w_V(i)| \leq 1$ induces a set $A'\subset V'$ in $G'$ with $A' = \bigcup_{i \in A} \{i\} \times [\lceil w_V(i) N \rceil]$, and $A'$ satisfies
\[
\left||A'| - |V' \setminus A'|\right| \EQ \left|\sum_{i \in A} \lceil w_V(i) N \rceil - \sum_{i \not\in A} \lceil w_V(i) N \rceil\right| \LE N + |V|\;.
\]
Since $|V'| \geq \sum_{i \in V}\mu(i)N \geq \eta|V|N$, the above is less than $2\eps |V|'$ as long as $N \geq |V| \geq 1/(\eps \eta)$. It follows that $A'$ achieves the same cut fraction as $A$ up to a factor of $(1 - O(\eps))$.
On the other hand, if we have some bisection $A' \subseteq V'$, then it induces the probabilities $p: V \to [0, 1], v \mapsto \frac{|A' \cap (\{i\} \times [\lceil \mu(i) N \rceil])|}{\lceil \mu(i) N \rceil}$. By paying $O(\delta)$ in the approximation ratio, we may move the vertices in $A'$ and assume that $\delta \leq p(i) \leq 1 - \delta$ for every $i \in V$. Now consider an set of independent Bernoulli random variables $\{X_i \mid i \in V\}$ where $\Pr[X_i = 1] = p_i$. Since each $\mu(i)$ lies in $[\eta, 1]$, we have that with high probability $\sum_{i \in V} X_i\mu(i) = (1/2 + o(1))\sum_{i \in V}p_i\mu(i)$. Now let $A$ be the (random) set of vertices such that $X_i = 1$, then $A$ is a near-bisection with high probability and by Lemma~\ref{lem:expander_mixing}, the cut value that it achieves is the same as that of $A'$, up to a multiplicative factor of $(1 \pm \eps)$. By choosing $\eps, \delta$ depending on $\tau$ appropriately, we obtain an $(\alpha + \tau)$-approximation.
\end{proof}

We finish the proof of our main theorem by using the same reduction as in Theorem~\ref{thm:weighted_unweighted_reduction}.

\begin{proof}[Proof of Theorem~\ref{thm:main}]
    We take the vertex-weighted, edge-weighted graph $G = (V, E, w_V, w_E)$ in Theorem~\ref{thm:main_thm_weighted} with parameters $\eps'=t'=O(\eps)$ and construct an unweighted graph $G' = (V', E')$ as in Theorem~\ref{thm:weighted_unweighted_reduction}. Note that in Theorem~\ref{thm:weighted_unweighted_reduction} in order to make the construction polynomial time we chose $N$ that is polynomially bounded in the size of $G$, but here we only consider existence, and in particular, we will make $N$ sufficiently large so we can for simplicity assume $w_V(i) N$ is an integer for every $i$ (we will also scale up $w_E$ to be large integers). Let $\bv_0, \bv_1, \ldots, \bv_n$ be an $\eps$-uncorrelated MAX BISECTION vector solution for $G$ as guaranteed by Theorem~\ref{thm:main_thm_weighted}. For any vertex $(i, k) \in \{i\} \times [w_V(i) N ] \subseteq V'$, we assign the unit vector $\bv_{i, k} = \bv_i$. Then the vectors $\bv_0, (\bv_{i, k})_{(i, k)\in V'}$ form an $\eps$-uncorrelated MAX BISECTION vector solution for $G'$, since
    \begin{equation}
        \sum_{(i, k) \in V'} \bv_0 \cdot \bv_{i, k} \EQ \sum_{i \in V}w_V(i) N \bv_0 \cdot \bv_i \EQ 0\;,
    \end{equation}
    and
    \begin{equation}
       \frac{\sum_{(i_1, k_1), (i_2, k_2) \in V'}  |\bv_{i_1, k_1}^\perp \cdot \bv_{i_2, k_2}^\perp|}{(\sum_{i \in V}w_V(i)N)^2} \EQ \frac{\sum_{i_1, i_2\in V}w_V(i_1)w_V(i_2)|\bv_{i_1}^\perp \cdot \bv_{i_2}^\perp|}{(\sum_{i \in V}w_V(i))^2}\;.
    \end{equation}
    By the way we constructed $E'$, it follows that triangle inequalities are also preserved on every edge in $E'$, and the completeness is preserved. By the analysis of Theorem~\ref{thm:weighted_unweighted_reduction}, the soundness is also preserved up to $O(\eps)$.
\end{proof}

\section{Conclusion}\label{sec:conclusion}

In this paper, we presented evidence that the optimal approximation ratio for MAX BISECTION might be strictly less than the Goemans-Williamson approximation ratio for MAX CUT. More precisely, we constructed a sequence of graphs $G_{\eps}$ for MAX BISECTION for which the ratio between $\OPT(G_{\eps})$ and the value of the optimal $\eps$-unbiased vector solution is less than $0.87853$. %
We conclude the paper with a few directions for future research.

\begin{enumerate}
\item The primary open question is finding a conditional NP-hardness proof that the approximation ratio for MAX BISECTION is less than $0.87853$. Some progress is being made to produce hardness results assuming the Small Set Expansion Hypothesis (SSEH)~\cite{GhLe23}, but the current loss in the arguments is too great to leverage our integrality gap.

\item Short of conditional hardness, another question would be to improve the integrality gap instance. For example, one could impose triangle inequalities on all pairs of vertices rather than just the ones for which there is an edge, although as Raghavendra~\cite{R08} shows, only the former is needed for UGC hardness of MAX CSPs. One could also more ambitiously hope to construct SoS integrality gaps, although analogous ones even for MAX CUT are a long-standing open problem. %

\item Another direction is to identify similar integrality gaps/dictatorship tests for other 2-variable CSPs. As previously discussed, the approximation ratio for MAX BISECTION 2-SAT is known to equal the approximation ratio for MAX 2-SAT~\cite{LLZ02,Austrin07,ABG16,BHZ24}. Does a similar phenomenon happen for, say, MAX BISECTION DI-CUT? We predict the answer is no, as the currently best-known $\THRESH$ rounding scheme for MAX DI-CUT~\cite{BHPZ23} is a non-linear, and also a non-monotone, function of the biases.

\item On the algorithmic side, one drawback of the Raghavendra-Tan/Austrin-Benabbas-Georgiou paradigm is that to compute an $\eps$-unbiased set of SDP vectors, we need $n^{O(1/\eps^4)}$ time. Is it possible to compute these vectors with fewer rounds of SoS or some other faster algortihm? We remark that the trade-off between the number of rounds and the average covariance is known to be tight in Raghavendra-Tan's correlation rounding~\cite{jain2019mean}.

\item Finally, there remains the possibility that a different rounding for higher levels of the SoS hierarchy might achieve $\agw$ or ($\agw - \eps$ for every $\eps > 0$ as conjectured in~\cite{ABG16}) for MAX BISECTION. However, this would seem to require completely new ideas.

\end{enumerate}

\section*{Acknowledgments}

We thank Venkatesan Guruswami for encouraging us to publish our investigation of MAX BISECTION.

\appendix

\section{Interval Arithmetic Verification}\label{appendix:ia}
In this appendix, we describe some details for verifying the upper bound on $\sound(\mathcal{D}^*)$ using interval arithmetic.\footnote{The verification code is available at \url{https://github.com/jbrakensiek/MAX-BISECTION}.} The interval arithmetic algorithm was implemented in C++ building on the Arb/FLINT interval arithmetic library \cite{johansson2017arb,johansson2018numerical,johansson2019computing}. More precisely, we implemented our algorithm using C++ wrappers for Arb specifically designed for computing the bivariate normal CDFs which are needed to estimate the soundness of blueprints~\cite{BHPZ23,BHZ24}. %

Our overall verification strategy is similar to the verification of MAX DI-CUT approximation ratio in~\cite{BHPZ23}, namely we use a recursive divide-and-conquer routine: we divide the parameter space into regions and verify each of them, and if for some region the result is inconclusive, then we subdivide that region further and verify recursively. The main new ingredient in our verification here is a binary search procedure with intervals. Recall that we reduced a 4-parameter problem $s(t_1, t_2, t_3, t_5)$ (see the proof of Lemma~\ref{lem:hard_dist} for definition of this function) to a two-parameter problem $s(t_1, t_2)$. In doing so, we need to find $t_3$ which is the root of a monotone function $\frac{\partial}{\partial t_3}s(t_1, t_2, t_3, t_5)$ for some given $t_1$ and $t_2$ ($t_5$ does not appear in the partial derivative.). Without interval arithmetic, this can be done via any standard root-finding procedure. However, with interval arithmetic, each evaluation gives us an interval which complicates the computation. To circumvent this, we find the root using the following simple variant on binary search.
\begin{algorithm}[H]
\small
    \caption{The root finding procedure for the partial derivative}\label{alg:root_finding}
\begin{algorithmic}[1]
\Procedure{RootLowerBound}{$t_1, t_2, \eps$} \Comment{$\eps$ is the accuracy parameter, }
\State \Comment{which may be chosen to be the average length of $t_1$ and $t_2$}
    \State $L \gets -1$, $R \gets 1$
    \While{$R - L \geq \eps$}
        \State $M \gets (R - L) / 2$
        \If {$\textproc{Partial}(t_1, t_2, M) > 0$} \Comment{Evaluates the partial derivative on given intervals}
            \State $L \gets M$
        \Else 
            \State $R \gets M$
        \EndIf
    \EndWhile
    \State \Return $\textproc{Join}(L, R)$ \Comment{Returns an interval that contains both $L$ and $R$}
\EndProcedure

\Procedure{RootUpperBound}{$t_1, t_2, \eps$}
    \State $L \gets -1$, $R \gets 1$
    \While{$R - L \geq \eps$}
        \State $M \gets (R - L) / 2$
        \If {$\textproc{Partial}(t_1, t_2, M) < 0$}
            \State $R \gets M$
        \Else 
            \State $L \gets M$
        \EndIf
    \EndWhile
    \State \Return $\textproc{Join}(L, R)$
\EndProcedure

\Procedure{RootFinding}{$t_1, t_2, \eps$}
    \State $L \gets \textproc{RootLowerBound}(t_1, t_2)$
    \State $U \gets \textproc{RootUpperBound}(t_1, t_2)$
    \State \Return $\textproc{Join}(L, U)$
\EndProcedure
\end{algorithmic}
\end{algorithm}

In words, we find both a lower bound and an upper bound for the root, and return an interval with them being the endpoints. To find a lower bound, we use the standard binary search procedure: we check at each iteration whether the partial derivative is strictly greater than 0, if so, then this constitutes a proof that the root lies to the right of the current point (recall that the partial derivative monotonically decreases); otherwise, we may not gain any information on where the root is since the evaluation may return an interval that contains 0 in its interior, but in this case we can choose to err on the safe side and keep searching on the left, since we only ask for a lower bound on the root. We proceed analogously the same for the upper bound.

\section{Rough Theoretical Analysis}\label{app:theory}

In this appendix, we give a rough theoretical analysis showing that the approximation ratio for MAX BISECTION where we have negated variables (i.e., MAX 2-LIN with the bisection constraint) is strictly less than $\alpha_{GW}$ (By allowing negated variables, we may assume that the threshold function is odd.).  To do this, we first give an approximation for the performance of threshold functions on configurations $(b_1,b_2,\bgw)$ where $b_1$ and $b_2$ are small. We then consider the following configurations
\begin{enumerate}
\item $(b,0,b_{GW})$
\item $(-b,2b,b_{GW})$
\item $(2b,-2b,b_{GW})$
\end{enumerate}
where $b > 0$ is a sufficiently small constant and show that if we choose appropriate weights for these configurations, any odd rounding scheme has performance at most $\alpha_{GW} - \Omega(b^4)$.

\begin{definition}
We define $\Phi(c) = \frac{1}{\sqrt{2\pi}}\int_{-\infty}^{c}{e^{-\frac{x^2}{2}}dx}$
\end{definition}
\begin{lemma}
When $c \approx 0$, $\int_{0}^{c}{e^{-\frac{x^2}{2}}dx} \approx c - \frac{c^3}{6}$
\end{lemma}
\begin{proof}
This follows by observing that $e^{-\frac{x^2}{2}} = 1 - \frac{x^2}{2} + O(x^4)$ and integrating.
\end{proof}
\begin{corollary}
When $c \approx 0$, $\Phi(c) \approx \frac{1}{2} + \frac{c}{\sqrt{2\pi}} - \frac{c^3}{6\sqrt{2\pi}}$.
\end{corollary}
\begin{definition}
We define $\Phi_{\rho}(c_1,c_2) = \frac{1}{2{\pi}\sqrt{1 - {\rho}^2}}\int_{-\infty}^{c_1}{\int_{-\infty}^{c_2}{e^{-\frac{x_1^2 - 2{\rho}{x_1}x_2 + x_2^2}{2(1 - {\rho}^2)}}d{x_2}}d{x_1}}$. 
\end{definition}%
\begin{remark}
This definition is different than our previous definition as we are taking the input for $\Gamma$ to be the limits of the integrals rather than the target probabilities for each variable being rounded to $1$.
\end{remark}
\begin{lemma}
When $c_1,c_2 \approx 0$, 
\begin{align*}
\Phi_{\rho}(c_1,c_2) &\approx \frac{1}{2} - \frac{\cos^{-1}(\rho)}{2\pi} + \frac{c_1 + c_2}{2\sqrt{2\pi}} - \frac{{\rho}c_1^2 - 2{c_1}{c_2} + {\rho}c_2^2}{4\pi\sqrt{1 - {\rho}^2}} - \frac{c_1^3 + c_2^3}{12\sqrt{2\pi}} \\
&+ \frac{(3{\rho} - 2{\rho}^3){c_1^4} -4{c_1^3}{c_2} + 6{\rho}{c_1^2}{c_2^2} - 4{c_1}{c_2^3} + (3{\rho} - 2{\rho}^3){c_2^4}}{48{\pi}(1 - {\rho}^2)^{\frac{3}{2}}}
\end{align*}
\end{lemma}
\begin{proof}
We make the following observations:
\begin{enumerate}
\item $\frac{1}{2{\pi}\sqrt{1 - {\rho}^2}}\int_{-\infty}^{0}{\int_{-\infty}^{0}{e^{-\frac{x_1^2 - 2{\rho}{x_1}x_2 + x_2^2}{2(1 - {\rho}^2)}}d{x_2}}d{x_1}} = \frac{1}{2} - \frac{\cos^{-1}(\rho)}{2\pi}$
\item 
\begin{align*}
&\frac{1}{2{\pi}\sqrt{1 - {\rho}^2}}\int_{0}^{c_1}{\int_{-\infty}^{0}{e^{-\frac{x_1^2 - 2{\rho}{x_1}x_2 + x_2^2}{2(1 - {\rho}^2)}}d{x_2}}d{x_1}} = 
\frac{1}{2{\pi}}\int_{0}^{c_1}{e^{-\frac{x_1^2}{2}}\int_{-\infty}^{-\frac{{\rho}x_1}{\sqrt{1 - {\rho}^2}}}{e^{-\frac{u_2^2}{2}}d{u_2}}d{x_1}} \\
&=\frac{1}{2\sqrt{2\pi}}\int_{0}^{c_1}{e^{-\frac{x_1^2}{2}}d{x_1}} + \frac{1}{2{\pi}}\int_{0}^{c_1}{e^{-\frac{x_1^2}{2}}\int_{0}^{-\frac{{\rho}x_1}{\sqrt{1 - {\rho}^2}}}{e^{-\frac{u_2^2}{2}}d{u_2}}d{x_1}} \\
&\approx \frac{1}{2\sqrt{2\pi}}\left(c_1 - \frac{c_1^3}{6}\right) + \frac{1}{2\pi}\int_{0}^{c_1}{\left(1 - \frac{x_1^2}{2}\right)\left(-\frac{{\rho}x_1}{\sqrt{1 - {\rho}^2}} + \frac{{{\rho}^3}x_1^3}{6(1 - {\rho}^2)^{\frac{3}{2}}}\right)d{x_1}} \\
&\approx \frac{1}{2\sqrt{2\pi}}\left(c_1 - \frac{c_1^3}{6}\right) - \frac{1}{4\pi}\left(\frac{{\rho}c_1^2}{\sqrt{1 - {\rho}^2}} - \frac{3{\rho} - 2{\rho}^3}{12(1 - {\rho}^2)^{\frac{3}{2}}}c_1^4\right)
\end{align*}
\item By symmetry, 
\begin{align*}
&\frac{1}{2{\pi}\sqrt{1 - {\rho}^2}}\int_{-\infty}^{c_1}{\int_{0}^{c_2}{e^{-\frac{x_1^2 - 2{\rho}{x_1}x_2 + x_2^2}{2(1 - {\rho}^2)}}d{x_2}}d{x_1}} \\
&\approx \frac{1}{2\sqrt{2\pi}}\left(c_2 - \frac{c_2^3}{6}\right) - \frac{1}{4\pi}\left(\frac{{\rho}c_2^2}{\sqrt{1 - {\rho}^2}} - \frac{3{\rho} - 2{\rho}^3}{12(1 - {\rho}^2)^{\frac{3}{2}}}c_2^4\right)
\end{align*}
\item 
\begin{align*}
&\frac{1}{2{\pi}\sqrt{1 - {\rho}^2}}\int_{0}^{c_1}{\int_{0}^{c_2}{e^{-\frac{x_1^2 - 2{\rho}{x_1}x_2 + x_2^2}{2(1 - {\rho}^2)}}d{x_2}}d{x_1}} \\
&\approx \frac{1}{2{\pi}\sqrt{1 - {\rho}^2}}\int_{0}^{c_1}{\int_{0}^{c_2}{1-\frac{x_1^2 - 2{\rho}{x_1}x_2 + x_2^2}{4(1 - {\rho}^2)}d{x_2}}d{x_1}} \\
&\approx \frac{{c_1}{c_2}}{2{\pi}\sqrt{1 - {\rho}^2}} - \frac{2{c_1^3}{c_2} - 3{\rho}{c_1^2}{c_2^2} + 2{c_2}{c_1^3}}{24{\pi}(1 - {\rho}^2)^{\frac{3}{2}}}
\end{align*}
\end{enumerate}
The result follows by adding the contributions from these terms.
\end{proof}
\subsection{Fixing $b_{12} = {\bgw}$}
If we fix $b_{12} = {b_1}{b_2} + {\rho}\sqrt{(1-b_1^2)(1-b_2^2)} = {\bgw}$ then we have that 
\[
\rho = \frac{{\bgw} - {b_1}{b_2}}{\sqrt{(1-b_1^2)(1-b_2^2)}} \approx {\bgw}\left(1 + \frac{b_1^2 + b_2^2}{2} + \frac{3b_1^4 + 2{b_1^2}b_2^2 + 3b_2^4}{8}\right) - b_{1}b_{2} - \frac{{b_1^3}b_2 + b_1{b_2^3}}{2}
\]
Since $\cos^{-1}({\bgw} + x) \approx \cos^{-1}({\bgw}) - \frac{x}{\sqrt{1 - {\bgw}^2}} - \frac{{{\bgw}}x^2}{2(1 - {\bgw}^2)^{\frac{3}{2}}}$, we have that $\cos^{-1}(\rho)$ is approximately
\begin{align*}
&\cos^{-1}({\bgw}) - \frac{{\bgw}b_1^2 - 2{b_1}{b_2} + {\bgw}b_2^2}{2\sqrt{1 - {\bgw}^2}} - \frac{3{\bgw}b_1^4 -4{b_1^3}b_2 + 2{\bgw}{b_1^2}b_2^2 -4b_1{b_2^3} + 3{\bgw}b_2^4}{8\sqrt{1 - {\bgw}^2}} \\
&-\frac{{{\bgw}}\left({\bgw}b_1^2 - 2{b_1}{b_2} + {\bgw}b_2^2\right)^2}{8(1 - {\bgw}^2)^{\frac{3}{2}}}
\end{align*}
Since $\frac{1}{\sqrt{1 - ({\bgw} + x)^2}} \approx \frac{1}{\sqrt{1 - {\bgw}^2}} + \frac{{{\bgw}}x}{(1 - {\bgw}^2)^{\frac{3}{2}}}$, we have that 
\[
\frac{1}{\sqrt{1 - \rho^2}} \approx \frac{1}{\sqrt{1 - {\bgw}^2}} + \frac{{\bgw}\left({\bgw}b_1^2 - 2{b_1}{b_2} + {\bgw}b_2^2\right)}{2\left(1 - {\bgw}^2\right)^{\frac{3}{2}}}
\]
Putting everything together, 
\begin{align*}
\Phi_{\rho}(c_1,c_2) &\approx \frac{1}{2} - \frac{\cos^{-1}({\bgw})}{2\pi} + \frac{c_1 + c_2}{2\sqrt{2\pi}} + \frac{{\bgw}b_1^2 - 2{b_1}{b_2} + {\bgw}b_2^2}{4{\pi}\sqrt{1 - {\bgw}^2}} - \frac{{\bgw}c_1^2 - 2{c_1}{c_2} + {\bgw}c_2^2}{4\pi\sqrt{1 - {\bgw}^2}} \\
&- \frac{c_1^3 + c_2^3}{12\sqrt{2\pi}} + \frac{3{\bgw}b_1^4 -4{b_1^3}b_2 + 2{\bgw}{b_1^2}b_2^2 -4b_1{b_2^3} + 3{\bgw}b_2^4}{16{\pi}\sqrt{1 - {\bgw}^2}} \\
&+\frac{{{\bgw}}\left({\bgw}b_1^2 - 2{b_1}{b_2} + {\bgw}b_2^2\right)^2}{16{\pi}(1 - {\bgw}^2)^{\frac{3}{2}}} - \frac{\left({\bgw}b_1^2 - 2{b_1}{b_2} + {\bgw}b_2^2\right)(c_1^2 + c_2^2)}{8\pi\sqrt{1 - {\bgw}^2}}\\
&-\frac{{\bgw}\left({\bgw}b_1^2 - 2{b_1}{b_2} + {\bgw}b_2^2\right)({\bgw}c_1^2 - 2{c_1}{c_2} + {\bgw}c_2^2)}{8\pi\left(1 - {\bgw}^2\right)^{\frac{3}{2}}}\\
&+ \frac{(3{\bgw} - 2{\bgw}^3){c_1^4} -4{c_1^3}{c_2} + 6{\bgw}{c_1^2}{c_2^2} - 4{c_1}{c_2^3} + (3{\bgw} - 2{\bgw}^3){c_2^4}}{48{\pi}(1 - {\bgw}^2)^{\frac{3}{2}}}
\end{align*}
Recalling that the soundness is $\Phi(c_1) + \Phi(c_2) - 2\Phi_{\rho}(c_1,c_2)$ %
, we have that the soundness is approximately
\begin{align*}
&\frac{\cos^{-1}({\bgw})}{\pi} - \frac{{\bgw}b_1^2 - 2{b_1}{b_2} + {\bgw}b_2^2}{2{\pi}\sqrt{1 - {\bgw}^2}} + \frac{{\bgw}c_1^2 - 2{c_1}{c_2} + {\bgw}c_2^2}{2\pi\sqrt{1 - {\bgw}^2}}\\
&-\frac{(9{\bgw} - 6{\bgw}^3)b_1^4 - 12{b_1^3}b_2 + 18{\bgw}{b_1^2}{b_2^2} - 12b_1{b_2^3} + (9{\bgw} - 6{\bgw}^3)b_2^4}{24{\pi}(1 - {\bgw}^2)^{\frac{3}{2}}}\\
&+\frac{6{\bgw}(b_1^2 + b_2^2)(c_1^2 + c_2^2) - 12{\bgw}^2(b_1^2 + b_2^2){c_1}{c_2}-12{b_1}{b_2}(c_1^2 + c_2^2) + 24{\bgw}{b_1}{b_2}{c_1}{c_2}}{24{\pi}(1 - {\bgw}^2)^{\frac{3}{2}}} \\
&-\frac{(3{\bgw} - 2{\bgw}^3){c_1^4} -4{c_1^3}{c_2} + 6{\bgw}{c_1^2}{c_2^2} - 4{c_1}{c_2^3} + (3{\bgw} - 2{\bgw}^3){c_2^4}}{24{\pi}(1 - {\bgw}^2)^{\frac{3}{2}}}
\end{align*}

As a sanity check, we consider what happens when we use hyperplane rounding. When we use hyperplane rounding, we have that $c_1 = \frac{{g}b_1}{\sqrt{1 - b_1^2}} \approx {g}(b_1 + \frac{b_1^3}{2})$ and $c_2 = \frac{{g}b_2}{\sqrt{1 - b_2^2}} \approx {g}(b_2 + \frac{b_2^3}{2})$ where $g$ is a Gaussian variable. This gives that the soundness is approximately the expected value over $g$ of 
\begin{align*}
&\frac{\cos^{-1}({\bgw})}{\pi} + \frac{{\bgw}b_1^2 - 2{b_1}{b_2} + {\bgw}b_2^2}{2{\pi}\sqrt{1 - {\bgw}^2}}\left(g^2 - 1\right) \\
&+\frac{-(9{\bgw}- 6{\bgw}^3) + (18{\bgw} - 12{\bgw}^3)g^2 - (3{\bgw} - 2{\bgw}^3)g^4}{24{\pi}(1 - {\bgw}^2)^{\frac{3}{2}}}(b_1^4 + b_2^4)\\
&+\frac{12 -24g^2 + 4g^4}{24{\pi}(1 - {\bgw}^2)^{\frac{3}{2}}}({b_1^3}b_2 + {b_1}b_2^3) + \frac{-18{\bgw} + 36{\bgw}g^2 - 6{\bgw}g^4}{24{\pi}(1 - {\bgw}^2)^{\frac{3}{2}}}{b_1^2}{b_2^2}
\end{align*}
which is $\frac{\cos^{-1}({\bgw})}{\pi}$.

Roughly speaking, if we have $\frac{1}{3}$ weight on bias $b_2$, $\frac{2}{3}$ weight on bias $-b_1$, and $b_2 = 2b_1$ then the constraint that the cut is balanced gives that $c_2 \approx 2c_1$ (this is not an exact equality but the error is a lower order term).\footnote{To make this rigorous, we also need to estimate for $|c_1|$ and $|c_2|$ sufficiently large that the soundness drops well below $\cos^{-1}(\bgw)/\pi$.} We also have that if $b = 0$ then $c = 0$. Thus, if $c$ is the threshold for $b$ then we have the following performance for the three types of clauses
\begin{enumerate}
\item For clauses with $b_1 = b$ and $b_2 = 0$, the soundness is approximately
\[
\frac{\cos^{-1}({\bgw})}{\pi} - \frac{{\bgw}(b^2 - c^2)}{2{\pi}\sqrt{1 - {\bgw}^2}} -\frac{(3{\bgw} - 2{\bgw}^3)(3b^4 - c^4) - 6{\bgw}{b^2}c^2}{24{\pi}(1 - {\bgw}^2)^{\frac{3}{2}}}
\]
\item For clauses with $b_1 = 2b$ and $b_2 = -b$, the soundness is approximately
\begin{align*}
&\frac{\cos^{-1}({\bgw})}{\pi}-\frac{(5{\bgw} + 4)(b^2 - c^2)}{2{\pi}\sqrt{1 - {\bgw}^2}} \\
&-\frac{(40 + 75{\bgw} - 34{\bgw}^3)(3b^4 - c^4) - (120 + 246{\bgw} + 120{\bgw}^2){b^2}c^2}{24{\pi}(1 - {\bgw}^2)^{\frac{3}{2}}}
\end{align*}
\item For clauses with $b_1 = 2b$ and $b_2 = -2b$, the soundness is approximately
\begin{align*}
&\frac{\cos^{-1}({\bgw})}{\pi}-\frac{(8{\bgw} + 8)(b^2 - c^2)}{2{\pi}\sqrt{1 - {\bgw}^2}} \\
&-\frac{(128 + 192{\bgw} - 64{\bgw}^3)(3b^4 - c^4) - (384 + 768{\bgw} + 384{\bgw}^2){b^2}c^2}{24{\pi}(1 - {\bgw}^2)^{\frac{3}{2}}}
\end{align*}
\end{enumerate}
We now choose weights $w_1$, $w_2$, and $w_3$ for these three configurations so that the $(b^2 - c^2)$ and $(3b^4 - c^4)$ terms cancel which leaves a negative coefficient times ${b^2}{c^2}$. In particular, since ${\bgw} \approx -0.6891577$, if we take the first configuration with weight $w_1$, the second configuration with weight $w_2$, and the third configuration with weight $w_3$ then the performance will be approximately
\begin{align*}
&\frac{\cos^{-1}({\bgw})}{\pi} + (.151368w_1 - .121728w_2 - .546192w_3)(b^2 - c^2)\\
&+(.005672w_1 + .002241w_2 - .066760w_3)(3b^4 - c^4)\\
&-(.016600w_1 + .029948w_2 + 0.148955){b^2}{c^2}
\end{align*}
If we set $w_1 \approx 0.53777$, $w_2 \approx .40301$, and $w_3 \approx 0.05922$ then this is approximately $\frac{\cos^{-1}({\bgw})}{\pi}-.02982{b^2}c^2$. 

If we modify the coefficients to be $w'_1 = w_1 - \frac{.01b^2}{.546192} \approx 0.537773 - \frac{.01b^2}{.546192}$, $w'_2 = w_2 \approx .40301$, and $w'_3 = w_3 - \frac{.01b^2}{.546192} \approx 0.05922 - \frac{.01b^2}{.546192}$ then the performance will be at most $\frac{\cos^{-1}({\bgw})}{\pi}-.01{b^4}$.

\bibliographystyle{alpha}
\bibliography{ref.bib}

\end{document}